\documentclass[showpacs,preprintnumbers,amssymb,twocolumn]{revtex4-2}
\usepackage{graphicx}
\usepackage{dcolumn}
\usepackage{color}
\usepackage{bm}
\usepackage{amsmath}
\usepackage{amssymb}
\usepackage{epsfig}
\usepackage{amsfonts}
\usepackage{lineno,hyperref}
\usepackage{array}
\usepackage{float}
\usepackage{microtype}
\usepackage{multirow}
\usepackage{adjustbox}
\usepackage[english]{babel}
\usepackage{subfigure}
\usepackage{blindtext}
\usepackage[a4paper, total={7.5in, 10in}]{geometry}

\def \a{\alpha}
\def \b{\beta}
\def \l{\lambda}
\def \L{\Lambda}

\def \k{\kappa}

\def \be{\begin{equation}}
\def \ee{\end{equation}}
\def \ben{\begin{eqnarray}}
\def \een{\end{eqnarray}}

\def \O{\Omega}

\def \t{\theta}
\def \P{\Phi}
\def \M{\mathcal{M}}
\def \G{\bar{G}}

\def \k{\kappa}
\def \R{\bar{R}}
\def \T{\bar{T}}

\def \T{\bar{T}}

\def \La{\mathcal{L}}

\begin{document}
\title{Geodesic structure of generalized Vaidya spacetime through the K-essence}

\author{Bivash Majumder}
\altaffiliation{bivashmajumder@gmail.com}
\affiliation{Department of Mathematics, Prabhat Kumar College, Contai, Purba Medinipur 721404, India}

\author{Maxim Khlopov}
\altaffiliation{khlopov@apc.in2p3.fr}
\affiliation{Institute of Physics, Southern Federal University, 194 Stachki, Rostov-on-Don 344090, Russian Federation $\&$\\
National Research Nuclear University, MEPHI, Moscow, Russian Federation $\&$\\ Virtual Institute of Astroparticle Physics 10, rue Garreau, 75018 Paris, France}
 
\author{Saibal Ray}
\altaffiliation{saibal.ray@gla.ac.in}
\affiliation{Centre for Cosmology, Astrophysics and Space Science (CCASS), GLA University, Mathura 281406, Uttar Pradesh, India}

\author{Goutam Manna$^{b}$ }
\altaffiliation{goutammanna.pkc@gmail.com\\
$^{b}$Corresponding author}
\affiliation{Department of Physics, Prabhat Kumar College, Contai, Purba Medinipur 721404, West Bengal, India $\&$\\  Institute of Astronomy, Space and Earth Science (IASES), Kolkata 700054, West Bengal, India}

\date{Received: date / Accepted: date}  

\begin{abstract}
This article investigates on the radial and non-radial geodesic structures of the generalized K-essence Vaidya spacetime. Within the framework of K-essence geometry, it is important to note that the metric does not possess conformal equivalence to the conventional gravitational metric. This study employs a non-canonical action of the Dirac-Born-Infeld kind. In this work, we categorize the generalized K-essence Vaidya mass function into two distinct forms. Both the forms of the mass functions have been extensively utilized to analyze the radial and non-radial time-like or null geodesics in great details inside the comoving plane. Indications of the existence of wormhole can be noted during the extreme phases of spacetime, particularly in relation to black holes and white holes, which resemble the Einstein-Rosen bridge. In addition, we have also detected the distinctive indication of the quantum tunneling phenomenon around the central singularity.
\end{abstract}

\maketitle
     
\section{Introduction}
S. Chandrasekhar extensively analyzed the time-like and null-geodesic features of the Schwarzschild spacetime in his book \cite{chandra}. 
In addition, he has examined the orbital configurations of both the confined and unconfined trajectories using graphical representations.  In addition, the authors in \cite{cruz} have examined the geodesic structures of the Schwarzschild anti-de Sitter spacetime.   The researchers assessed both radial and non-radial paths for time-like and null geodesics.   Additionally, they have demonstrated that the geodesic structures of this black hole exhibit distinct forms of motion that are not permitted by the Schwarzschild spacetime.  The geometric framework of the Schwarzschild spacetime is also examined in \cite{berti}. The Jacobi metric for time-like geodesics in static spacetimes has been examined in the reference \cite{gibbons}.   They have demonstrated that the unrestricted movement of large particles in stationary spacetimes is determined by the geodesics of a Riemannian metric that depends on the particle's energy. This metric is similar to Jacobi's metric in classical dynamics.   When the mass of an object approaches zero, Jacobi's metric becomes identical to the Fermat or optical metric, which does not depend on energy.   In addition, they have provided a detailed account of the characteristics of the Jacobi metric pertaining to the motion of heavy particles beyond the event horizon of a Schwarzschild black hole.   The authors of \cite{chanda} derived the Jacobi metric for different stationary metrics and developed the Jacobi-Maupertuis metric for time-dependent metrics by using the Eisenhart-Duval lift \cite{eisenhart,duval}. The authors in \cite{gm1} have documented the remarkable characteristics of the time-like geodesic structure when dark energy is present in an emergent gravity framework, specifically for the Barriola-Vilenkin metric \cite{bv}. The K-essence emergent gravity metric is precisely correlated with the Barriola-Vilenkin (BV) metric for the Schwarzschild background, specifically for a certain form of K-essence scalar field \cite{gm2}. The researchers have analyzed the various paths that time-like geodesics can take in the presence of dark energy in the Barriola-Vilenkin spacetime \cite{gm1}, which is equivalent to the Schwarzschild spacetime in terms of its fundamental structure. However, the permissible ranges for the maximum and minimum distances from the central object are significantly distinct.   For a constant dark energy density, the orbits, both bound and unbound, are graphed.

In 1951, P. C. Vaidya proposed the first relativistic line element that properly represented the spacetime of a conceivable star \cite{Vaidya}.  It extended the specific solution of Schwarzschild by depicting the emission of radiation for a mass that is not in a static state.   The Schwarzschild solution describes the geometry of spacetime around a spherically symmetric, non-rotating, black object with a constant mass.   Therefore, it is clear that the model is incapable of accurately depicting spacetime outside the confines of a star.  The solution proposed by Vaidya \cite{Vaidya}, known as the Vaidya spacetime or the radiating Schwarzschild metric, was introduced as a possible explanatory framework.   The main distinction between the two metrics is that the Vaidya metric adds a time-dependent mass parameter, whereas the Schwarzschild metric uses a constant mass value. As a result, the spacetime in the Vaidya metric evolves with time. The Vaidya metric is primarily used to investigate gravitational collapse.   The occurrence of gravitational collapse is widely acknowledged in the disciplines of general relativity and astrophysics, as demonstrated by the research conducted by Joshi et al. ~\cite{Joshi1,Joshi2,Joshi3,Joshi4, Joshi5,Joshi6,Oppenheimer,Malafarina}.  It plays a vital role in understanding several astrophysical aspects of our cosmos.   The phenomenon of gravitational collapse provides useful insights into several elements of astronomy, including the evolution of structures, the features of stars, the genesis of black holes, and the construction of white dwarfs or neutron stars, among other events. Gravitational collapse refers to the phenomenon in which a star collapses as a result of its mass. The outcome of this collapse might vary depending on the exact beginning mass conditions, leading to distinct stages of collapse.   Papapetrou \cite{papa} was the first to demonstrate that the solution of a null dust fluid with spherical symmetry in gravitational collapse can lead to the creation of naked singularities. This statement presents a counterexample of the cosmic censorship hypothesis (CCH) as proposed by Penrose \cite{penrose}.   The authors in \cite{Joshi3,vert} have provided a detailed account of the causal paths that connect the singularities in the continuing Vaidya scenario.   Furthermore, a comprehensive classification of the non-spacelike geodesics that link the naked singularity in the past is presented, offering a rather thorough discussion of the restrictions involved.   It is subsequently demonstrated to be a robust curvature singularity in a more significant manner.

The Vaidya solution, as a generalization, encompasses all the established solutions of Einstein's field equations that include a mix of Type-I and Type-II matter fields \cite{vai1,vai2,vai3,vai4,vai5,jg}. The composition of this work is attributed to Husain \cite{husain}  and Wang $\&$ Wu \cite{wang}.  The extension of the Vaidya solution is sometimes referred to as the generalized Vaidya spacetime. The work performed in \cite{maombi} examines the gravitational collapse of the generalized Vaidya spacetime within the framework of the cosmic censorship theory.   They demonstrated that the categories of generalized Vaidya mass functions had emerged in the situation, suggesting the end of collapse with a locally visible central singularity.   The authors computed the magnitude of these singularities. A comprehensive mathematical framework was created to examine the requirements for the mass function for non-spacelike geodesics going towards the future to end at the singularity in the past.   Furthermore, they demonstrated that, when considering a certain generalized Vaidya mass function, the ultimate outcome of the collapse can be precisely defined as either a black hole or a naked singularity. The work by Patil \cite{patil} examines the phenomenon of gravitational collapse in higher dimensions within the context of the charged-Vaidya spacetime.   It has been demonstrated that singularities occur in a charged null fluid in a higher dimension. These singularities consistently lack any form of covering, hence contradicting the strong CCH.    This idea does not specifically pertain to weak cosmic censorship.   The Vaidya metric has received significant attention in scholarly research, with several major contributions to our comprehension of this subject. The authors of the study \cite{Coudray} examined the geometric properties of Vaidya's spacetime while considering a white hole that undergoes a decrease in mass. They found that the white hole can either stabilize and transform into a black hole within a limited or indefinite amount of time, or entirely evaporate. The researchers have focused specifically on the scenario of total evaporation over an indefinite period of time. They have successfully demonstrated the presence of an asymptotic light-like singularity in the conformal curvature, which connects both the past space-like singularity and the future time-like infinity.  Vertogradov \cite{vert1} conducted a study on the structure of the generalized Vaidya spacetime, specifically focusing on the case when the matter field of type-II follows the equation of state $P=\rho$. The findings of the study revealed the presence of an eternal naked singularity in this spacetime, which meets all energy conditions.  Once formed, the singularity will remain perpetually uncovered by the apparent horizon.   Nevertheless, the formation of the apparent horizon leads to the emergence of a white hole. Solanki et al. \cite{Solanki} have derived precise mathematical equations that describe the changes in the photon sphere and the angular radius of the shadow in a certain Vaidya spacetime.   The mass function $m(v)$ has been seen as a function of time that either increases or decreases linearly.   The initial scenario can function as a basic representation of a black hole that is accumulating matter, whereas the subsequent scenario can be seen as an illustration of a black hole that is emitting radiation, as theorized by Hawking.

In the realm of K-essence geometry, Manna et al. \cite{gm3} were the first to establish a link between K-essence geometry and Vaidya spacetime. They achieved this by introducing a new definition of the generalized Vaidya mass function, which directly depends on the kinetic energy of the K-essence scalar field. Subsequently, Manna et al. \cite{gm4} demonstrated that the K-essence emergent gravity metric bears a strong resemblance to the recently found generalized Vaidya metrics for the collapse of a null fluid. This similarity arises from the presence of a k-essence emergent mass function. Notably, Manna's analysis exclusively considers the K-essence scalar field as a function of either the advanced or the retarded time.  The recently developed K-essence model, known as the K-essence emergent Vaidya spacetime, has successfully met all the necessary energy conditions.   The presence of the centrally exposed singularity and the intensity and stability of the singularities in the K-essence emergent Vaidya metric yield intriguing results in their research. The evaporation of the dynamical horizon with the Hawking temperature in the K-essence Vaidya Schwarzschild spacetime was investigated by Manna et al. in \cite{gm5}.   This study uses the dynamical horizon equation to quantify the reduction in mass caused by Hawking radiation. Additionally, the tunneling formalism, namely the Hamilton-Jacobi technique, is utilized to compute the Hawking temperature.   In addition, Sawayama's revised explanation of the dynamical horizon \cite{Sawayama} is utilized to demonstrate that the results obtained differ from the conventional Vaidya spacetime geometry. The authors have established by analytical measures that the mass of the black hole, denoted as $m(v, r)$, in the K-essence emergent Schwarzschild-Vaidya spacetime, consistently decreases over time but does not fully evaporate.

The K-essence theory is a scalar field model that deviates from the canonical form. In this theory, the dominant energy component of the field is its kinetic energy, rather than its potential energy. This concept has been extensively studied by several researchers \cite{Visser,Babichev1,Babichev2,Vikman,Babichev3,Chimento1,Picon1,Scherrer,Chimento2,Picon2,Picon3}.  The distinctions between the K-essence theory employing a non-canonical Lagrangian and the relativistic field theories utilizing a canonical Lagrangian are found in the sophisticated dynamical solutions of the K-essence equation of motion. These solutions not only spontaneously violate Lorentz invariance but also alter the metric for the perturbations around them.  The disturbances propagate in the emergent or analogous curved spacetime, characterized by a metric distinct from the gravitational metric. The non-canonical Lagrangian may be expressed as $\La(X)=-V(\phi)F(X)$, where, $X =\frac{1}{2} g_{\mu\nu} \nabla^{\mu} \phi \nabla^{\nu} \phi$, $\phi$ is the K-essence scalar field, $V(\phi)$ is the potential term.  An alternate form of the Lagrangian, as described by Tian~\cite{Tian}, may be represented as $\La=[1+f(y)]X+[1+g(y)]V_{exp}$, where $V_{exp}=V_{0}~exp(-\lambda\phi)$, $V_{0}$ and $\lambda$ are constants, $y=X/V_{exp}$, and $f(y)$ and $g(y)$ are arbitrary functions. The functions $f(y)$ and $g(y)$ are unrestricted and can have any form. Furthermore, it is important to mention that there exist examples of K-essence theories that are not minimally linked, as mentioned in the Refs. \cite{Myrzakulov2, Sen, Chatterjee}.  Nevertheless, this article only addresses the minimally coupled K-essence theory, as investigated by the Refs. \cite{Visser,Babichev1,Babichev2,Vikman,Babichev3,Chimento1,Picon1,Scherrer,Chimento2,Picon2,Picon3}. In a general sense, the Lagrangian has the capacity to depend on any functions of $\phi$ and $X$.  The K-essence theory offers the benefit of circumventing both the fine-tuning problem and the coincidence problem \cite{Velten} of the current universe. Additionally, it generates the negative pressure required for the universe's acceleration only through the kinetic energy of the field. The kinetic term of the field dominates over the potential term.  The article \cite{Picon1} presents attractor solutions where the dynamics of the cosmos are governed by the scalar field of the models. During the radiation-dominated phase, the K-essence field mimicked the equation of state of radiation and had a constant ratio to the radiation density. The K-essence field was unable to replicate the dust-like equation of state (EoS) due to dynamical limits during the time dominated by dust. However, it rapidly reduced its energy value by many orders of magnitude and eventually reached a constant value. Subsequently, over a period approximately equivalent to the current age of the universe, the density of matter was diminished by the K-essence field, leading to the commencement of cosmic acceleration.  The equation of state (EoS) of the K-essence theory ultimately converges to a value within the range of 0 to -1. Although in theory, it has the potential to extend beyond $-1$.   Another intriguing aspect of the K-essence idea is its potential to generate a type of dark energy where the speed of sound is consistently slower than that of light.   This feature may mitigate the cosmic microwave background (CMB) disruptions on large angular scales \cite{Erickson,Dedeo,Bean}. In this specific situation, Manna et al. \cite{gm1,gm2,gm3,gm4,gm5,gm6,gm7,gm8,gm9,gm10,gm11} have developed a fascinating emergent gravity metric referred to as $\G_{\mu\nu}$.   This metric possesses distinct attributes in contrast to the standard gravitational metric $g_{\mu\nu}$ and is derived from the notions of the Dirac-Born-Infeld (DBI) type action, as outlined in the works \cite{Mukohyama, Born, Heisenberg, Dirac}.   Dirac et al. proposed a non-canonical Lagrangian in order to eliminate the infinite self-energy of the electron, as described in their work \cite{Dirac}. The specific reasons and objectives for selecting the non-canonical theory, such as the K-essence theory, may be found in the Refs.\cite{gm12, gm13}. The Planck collaborations' findings, as shown in Refs. \cite{Planck1, Planck2, Planck3}, have examined the empirical evidence supporting the concept of K-essence with a DBI-type non-canonical Lagrangian, along with other modified theories.   Furthermore, it has been noted that the K-essence theory may be applied in a model that combines dark energy and dark matter \cite{gm1, gm2, gm6, gm7, gm8, gm9, Scherrer}, as well as from a purely gravitational perspective \cite{gm3, gm4, gm5, gm10, gm11}.\\

This article is organized as follows:
In section 2, we provided a concise explanation of the K-essence geometry and its connection to the conventional generalized Vaidya spacetime, which leads to the construction of a new generalized K-essence Vaidya spacetime. Section 3 offers a comprehensive analysis of the geodesic structures observed in the generalized K-essence Vaidya spacetime. This analysis considers two forms of mass function while ensuring that the condition on the kinetic energy of the K-essence scalar field is maintained. This section also provides a detailed analysis of the radial and non-radial geodesics used to examine the structure of time-like and null geodesics in the given spacetime. This is achieved by solving the Euler-Lagrange equations. The graphical and numerical analysis is also done in this section. In section 4, we will wrap up both the conclusion and the discussion.\\

\section{Brief of the relation between K-essence with generalized Vaidya spacetime}
This section offers a short introduction to the geometry of K-essence and the generalized Vaidya spacetime.   Initially, we present a brief summary of the geometric aspects related to the K-essence, as extensively explored in many scholarly references \cite{Visser,Babichev1,Babichev2,Vikman,Babichev3,Chimento1,Picon1,Scherrer,Chimento2,Picon2,Picon3}.   The action performed by this geometry is  
\ben
S_{k}[\phi,g_{\mu\nu}]= \int d^{4}x {\sqrt -g} \La(X,\phi)
\label{1}
\een
where the expression $X=\frac{1}{2}g^{\mu\nu}\nabla_{\mu}\phi\nabla_{\nu}\phi$ represents the canonical kinetic term, whereas $\La(X,\phi)$ denotes the non-canonical Lagrangian.  In this scenario, the conventional gravitational metric $g_{\mu\nu}$ has formed a minimum coupling with the K-essence scalar field ($\phi$). 

The energy-momentum tensor that corresponds solely to the K-essence scalar field is:
\ben
T_{\mu\nu}\equiv \frac{-2}{\sqrt {-g}}\frac{\delta S_{k}}{\delta g^{\mu\nu}}=-2\frac{\partial \La}{\partial g^{\mu\nu}}+g_{\mu\nu}\La\nonumber\\
=-\La_{X}\nabla_{\mu}\phi\nabla_{\nu}\phi
+g_{\mu\nu}\La,
\label{2}
\een
where $\La_{\mathrm X}= \frac{d\La}{dX},~ \La_{\mathrm XX}= \frac{d^{2}\La}{dX^{2}},
~\La_{\mathrm\phi}=\frac{d\La}{d\phi}$ and  $\nabla_{\mu}$ is the covariant derivative defined with respect to the gravitational metric $g_{\mu\nu}$. 

The equation of motion (EOM) for the K-essence scalar field is
\ben
-\frac{1}{\sqrt {-g}}\frac{\delta S_{k}}{\delta \phi}= \tilde{G}^{\mu\nu}\nabla_{\mu}\nabla_{\nu}\phi +2X\La_{X\phi}-\La_{\phi}=0,
\label{3}
\een
where  
\ben
\tilde{G}^{\mu\nu}\equiv \frac{c_{s}}{\La_{X}^{2}}[\La_{X} g^{\mu\nu} + \La_{XX} \nabla ^{\mu}\phi\nabla^{\nu}\phi],
\label{4}
\een
with $1+ \frac{2X  \La_{XX}}{\La_{X}} > 0$ and $c_s^{2}(X,\phi)\equiv{(1+2X\frac{\La_{XX}}
{\La_{X}})^{-1}}$.

Following \cite{gm1,gm2,gm3,gm4,gm6}, the inverse metric can be written as 
\ben
\bar{G}_{\mu\nu}=g_{\mu\nu}-\frac{\La_{XX}}{\La_{X}+2X\La_{XX}}\nabla_{\mu}\phi\nabla_{\nu}\phi.
\label{5}
\een
The Eqs. (\ref{4}), (\ref{5}) have physical relevance when $\La_{X}$ is nonzero, assuming a positive definite $c_{s}^{2}$.  Eq. (\ref{5}) states that the emergent metric, represented as $\bar{G}_{\mu\nu}$, differs in its conformal properties from the metric  $g _{\mu\nu}$ when considering non-trivial configurations of the scalar field $\phi$.  Like canonical scalar fields, the variable $\phi$ exhibits diverse local causal structural properties.   It also differs from those that are defined using $g_{\mu\nu}$.   The EOM, as stated in Eq. (\ref{3}), is valid even when taking into account the implicit relationship between $L$ and $\phi$. Then the EOM Eq. (\ref{3}) is:
\ben
\frac{1}{\sqrt{-g}}\frac{\delta S_{k}}{\delta \phi}= \bar G^{\mu\nu}\nabla_{\mu}\nabla_{\nu}\phi=0.
\label{6}
\een

This study addresses the Dirac-Born-Infeld (DBI) type non-canonical Lagrangian, which is represented as $\La(X,\phi)\equiv \La(X)$  \cite{gm1,gm2,gm3,Mukohyama,Born,Heisenberg,Dirac,gm12,gm13}:
\ben
\La(X)= 1-\sqrt{1-2X}.
\label{7}
\een

The K-essence paradigm posits that the prevalence of kinetic energy over potential energy results in the exclusion of the potential term in the Lagrangian equation (\ref{7}) \cite{Mukohyama,gm12,gm13}.   The squared speed of sound, represented as $c_{s}^{2}$, is determined by the expression $(1-2X)$.   Therefore, the Eq. (\ref{5}) for the effective emergent metric is expressed as:
\ben
\bar G_{\mu\nu}= g_{\mu\nu} - \nabla_{\mu}\phi\nabla_{\nu}\phi= g_{\mu\nu} - \partial_{\mu}\phi\partial_{\nu}\phi,
\label{8}
\een
since $\phi$ is a scalar. 

The Christoffel symbol corresponding to the emergent gravity metric given by Eq. (\ref{8}) can be written as \cite{gm1,gm2,gm12,gm13}: 
\ben
\bar\Gamma ^{\alpha}_{\mu\nu} 
&&=\Gamma ^{\alpha}_{\mu\nu} -\frac {1}{2(1-2X)}\Big[\delta^{\alpha}_{\mu}\partial_{\nu}
+ \delta^{\alpha}_{\nu}\partial_{\mu}\Big]X~~~~~~~~~~~
\label{9}
\een
where $\Gamma ^{\alpha}_{\mu\nu}$ is the usual Christoffel symbol associated with the gravitational metric $g_{\mu\nu}$.

Hence, the geodesic equation governing the K-essence geometry may be expressed as:
\ben
\frac {d^{2}x^{\alpha}}{d\l^{2}} +  \bar\Gamma ^{\alpha}_{\mu\nu}\frac {dx^{\mu}}{d\l}\frac {dx^{\nu}}{d\l}=0, \label{10}
\een
where $\l$ is an affine parameter.

The covariant derivative $D_{\mu}$ \cite{Babichev1,gm12,gm13} linked with the emergent metric $\bar{G}_ {\mu\nu}$ $(D_{\a}\bar{G}^{\a\b}=0)$ gives
\ben
D_{\mu}A_{\nu}=\partial_{\mu} A_{\nu}-\bar \Gamma^{\l}_{\mu\nu}A_{\l}, \label{11}
\een
and the inverse emergent metric is $\bar G^{\mu\nu}$ such as $\bar G_{\mu\l}\bar G^{\l\nu}=\delta^{\nu}_{\mu}$.

Therefore, considering the extensive behavior that defines the dynamics of K-essence and general relativity \cite{Vikman,gm12,gm13}, the Emergent Einstein's Equation (EEE) may be formulated as:

\ben
\mathcal{\G}_{\mu\nu}=\R_{\mu\nu}-\frac{1}{2}\bar{G}_{\mu\nu}\R=\k \T_{\mu\nu}, \label{12}
\een
where $\k=8\pi G$ is constant, $\R_{\mu\nu}$ is Ricci tensor and $\R~ (=\R_{\mu\nu}\bar{G}^{\mu\nu})$ is the Ricci scalar. Moreover, the energy-momentum tensor $\T_{\mu\nu}$ is linked to this emergent spacetime. 

Now, we will provide a concise overview of the K-essence emergent generalized Vaidya spacetime. 
In the cited work \cite{gm4}, the author has introduced the concept of K-essence emergent generalized Vaidya spacetime. This framework considers the background gravitational metric to be the typical generalized Vaidya metric \cite{husain, wang}, while also satisfying the necessary energy requirements.  The line element for the emergent generalized Vaidya metric in K-essence theory is as follows: 
\ben
dS^{2}&&=-\Big[1-\frac{2m(t,r)}{r}-\phi_{t}^{2}\Big]dt^{2}+2 dtdr+r^{2}d\O^{2}\nonumber\\
&&=-\Big[1-\frac{2\mathcal{M}(t,r)}{r}\Big]dt^{2}+2 dtdr+r^{2}d\O^{2}
\label{13}
\een
with $d\O^{2}=d\t^{2}+d\P^{2}$.
They have defined the K-essence emergent Vaidya mass function 
\ben
\mathcal{M}(t,r)=m(t,r)+\frac{r}{2}\phi_{t}^{2}
\label{14}
\een
where $m(t,r)$ is the usual generalized Vaidya mass function and $\phi_{t}^{2}$ ($\phi_{t}=\frac{\partial\phi}{\partial t}$) is the non-zero kinetic energy of the K-essence scalar field.
The above-mentioned mass function pertains to the gravitational energy associated with the K-essence emergent gravity within a specified radius $r$.   Here, we substitute the Edington advanced time coordinate with the conventional time coordinate, without any loss of generality, denoted as $v\rightarrow t$.   In this study \cite{gm4}, the author has examined the effective K-essence emergent metric, as denoted by Eq. (\ref{8}). Additionally, the author has calculated all the components of the EEE (Eq. (\ref{12})) and the necessary energy conditions. It is important to mention that the assumption about $\phi$ contradicts local Lorentz invariance since, in general, spherical symmetry only requires $\phi(x)=\phi(t,r)$.    The inclusion of the assumption of the independence of $\phi$, denoted as $\phi(t,r)=\phi(t)$, suggests that beyond this specific frame selection, a spherically symmetric $\phi$ is indeed a function of both $t$ and $r$.   The K-essence theory permits the occurrence of Lorentz violation due to the fact that the dynamic solutions of the K-essence equation of motion spontaneously break Lorentz invariance and alter the metric for the perturbations around these solutions.  

Furthermore, the authors in \cite{gm3} have successfully established a connection between the geometry of K-essence and the Vaidya spacetime.   The researchers have developed a model of the Vaidya spacetime with generalized K-essence, which takes into account any spherically symmetric static black hole as the underlying spacetime. The line element of the new geometry is (using Eq. (\ref{8}):
\ben
dS^2&&=-\Big[f(r)-\phi_{t}^{2}\Big]dt^{2}+2 dtdr+r^{2}d\O^{2}\nonumber\\
&&=-\Big(1-\frac{2\mathcal{M}(t,r)}{r} \Big)dt^2 + 2dtdr +r^2 d\O^2
\label{15} 
\een
gives the mass function 
\ben
\mathcal{M}(t,r) = \frac12 r\Big[1 + \phi_t^2 - f(r) \Big].
\label{16}
\een
In this article \cite{gm3}, the authors also have calculated all the components of EEE and required energy conditions.
If we consider $f(r) = (1-2M/r)$, i.e., the background physical spacetime is Schwarzschild spacetime, the mass function may be expressed as:
\ben
\mathcal{M}(t,r)=M+\frac{r}{2}\phi_t^2.
\label{17}
\een
Again, if we select the function $f(r)=1-\frac{2M}{r}+\frac{Q^{2}}{r^{2}}$, $Q$ represents the charge of the Reissner-Nordstrom (RN) black hole in the physical spacetime. In this case, the related mass function \cite{gm3} is modified as
\ben
\mathcal{M}(t,r)=M-\frac{Q^{2}}{2r}+\frac{r}{2}\phi_{t}^{2}.
\label{18}
\een
It is important to mention that the values of $\phi_{t}^{2}$ must be between 0 and 1. Otherwise, the metric (\ref{13}) and (\ref{15}) cannot be specified properly, and the presence of a dynamical horizon is also questionable \cite{gm3, gm4}.   In order to maintain the energy conditions, it is evident that the $\phi_t^{2}$ must be a monotonically increasing function of $t$, with the condition $\phi_t^2 < 1$.  The admissible configurations for the K-essence scalar field in the generalized Vaidya solution, in order to have a dynamical horizon, are subject to a highly restrictive constraint.  It is important to note that the metrics mentioned above represent dynamical horizons rather than isolated or event horizons, as explained extensively in Ref. \cite{gm3,gm4}.

It is also noted that the time dependence in the given mass functions (Eqs. (\ref{17}), (\ref{18})) arises from the kinetic energy of the K-essence scalar field. However, in the mass function Eq. (\ref{14}), the time dependence comes from both the usual generalized Vaidya mass and the K-essence scalar fields. Thus, considering the above situations of the K-essence generalized Vaidya spacetime, we may conclude that the K-essence Vaidya mass function adheres to the general form specified in Eq.(\ref{14}).   Therefore, we may conclude that the background metric can be chosen from any standard gravitational metric, with the only alteration being the replacement of their masses with a background mass, which likewise satisfies the EEE equation.

\section{Geodesics for the generalized K-essence Vaidya spacetime}
This section focuses on analyzing the geodesic structure of the generalized K-essence Vaidya spacetime.   In this context, we define our investigative metric as Eq. (\ref{13}), where the K-essence emergent Vaidya mass function is represented by Eq. (\ref{14}). For the metric (\ref{13}), we can write the  Lagrangian as \cite{chandra, gm1, vert1, Solanki}
\begin{equation}
    2\mathcal{L}=-\Big(1-\frac{2\mathcal{M}(t,r)}{r}\Big)\dot{t}^{2}+2\dot{t}\dot{r}+r^{2}\dot{\theta}^{2}+r^{2}\sin^{2}{\theta}~\dot{\Phi}^{2}.
\label{19}
\end{equation}
where
 $\dot t = \frac{dt}{d\tau},~\dot r = \frac{dr}{d\tau},~
\dot \t = \frac{d\t}{d\tau},~\dot \P = \frac{d\P}{d\tau}$, $\tau $ is to be identified with the proper time.
Now, using the Euler-Lagrange equation we have 
\begin{align}
    &\frac{d}{d\tau}\Big( \frac{\partial \mathcal{L}}{\partial  \dot{t}} \Big)=\frac{\mathcal{M}_{t}}{r}\dot{t}^{2};\label{20}\\
    &  2r\dot{\theta}+r^{2} \ddot{\theta}= r^{2}\sin{\theta}\cos{\theta}~\dot{\Phi}^{2}\label{21}\\
    and~~~& r^{2}\sin^{2}{\theta}~\dot{\Phi} =Constant\label{22}\\
    with ~~~& \frac{\partial \mathcal{L}}{\partial  \dot{t}}=-\Big(1-\frac{2\mathcal{M}}{r}\Big)\dot{t}+\dot{r} \label{23}
\end{align}
where $\M_{t}=\frac{\partial \M(t,r)}{\partial t}$ and we write $\M(t,r)$ as $\M$.

Because our object and metric are spherically symmetric, we can simplify everything by examining just motion on the equatorial plane $\theta=\frac{\pi}{2}$ and therefore, $\dot{\theta}=0$.
For the above choice of equatorial plane, the Eq. (\ref{22}) becomes
\ben
r^{2}\dot{\Phi}=Constant=L (say)
\label{24}
\een
Thus, by employing Eq. (\ref{24}) on the equatorial plane, we may write from Eq. (\ref{19}) that the Lagrangian is
\ben
    2\mathcal{L}=-\Big(1-\frac{2\mathcal{M}(t,r)}{r}\Big)\dot{t}^{2}+2\dot{t}\dot{r}+\frac{L^{2}}{r^{2}}.
\label{25}
\een

Due to the inclusion of the generalized K-essence Vaidya mass function ($\M(t,r)$) in the Lagrangian formulation provided above, further analysis is not possible as it can have varying values based on the gravitational mass.   Within this particular situation, we have the option to select the mass function.   For our subsequent analysis, we have selected two distinct mass functions. Moreover, it is mentioned that the K-essence Vaidya mass function (\ref{14}) depends on $\phi_{t}^{2}$, which has values between $0$ and $1$. Therefore, we can select $\phi_{t}^{2}$ as an explicit function of time in order to keep the values of $\phi_{t}^{2}$ throughout the article as \cite{gm3} 
\ben
\phi_{t}^{2}=e^{-t/t_{0}}
\label{26}
\een
where $t_{0}$ is a positive constant.

\subsection{Case-I: $\mathcal{M}(t,r)=M+\frac{r}{2}\phi_{t}^{2}$}
In this portion, we will look at the K-essence Vaidaya mass function as
\ben
\mathcal{M}(t,r)=M+\frac{r}{2}e^{-t/t_{0}}
\label{27}
\een
where $M$ is the Schwarzschild black hole's mass. In this scenario, the time dependence of the mass parameter is derived from the K-essence scalar field via Eq. (\ref{26}), which was previously discussed in the preceding section. Given that the mass parameter in Eq. (\ref{25}) is directly influenced by time through equations (\ref{14}) and (\ref{26}), we may infer from Refs. \cite{vert1, Solanki} that the energy $E$ can be expressed as a function of time $t$:
\ben
\frac{\partial \mathcal{L}}{\partial  \dot{t}}=E(t).
\label{28}
\een

Now using Eqs. (\ref{27}) and (\ref{28}) in Eq. (\ref{20}), we have
\ben
\dot{E}(t)=-\frac{e^{-\frac{t}{t_{0}}}}{2t_{0}}~\dot{t}^{2}
\label{29} 
\een

The solution of the aforementioned equation, as denoted by Eq. (\ref{29}), is highly intricate and cannot be solved directly. To determine the expression for $E(t)$ in the above equation, we converted our measurement to a comoving plane, where $\frac{dt}{d\tau}=1$. This conversion is consistently maintained throughout the article. Thus, on the comoving plane ($d\tau\equiv dt$), the expression for $E(t)$ is 
\ben
E(t)=\frac{1}{2} e^{-\frac{t}{t_{0}}}\equiv \frac{1}{2}\phi_{t}^{2}. \label{30}
\een

Thus, we can say that in our model, specifically when we select a specific form (\ref{26}) for the kinetic energy of the K-essence scalar field while satisfying the imposed conditions, we have found a direct relationship between the energy of the system we have chosen and the kinetic energy of the K-essence scalar field in the comoving plane. So, the K-essence Vaidya mass function (\ref{27}) can be written as
\ben
\mathcal{M}(t.r)=M+E(t)r. 
\label{31}
\een

It is important to point out that the Vaidya metric defines the gravitational field surrounding a massive object, often a dying star, that emits radiation in the form of null dust.  The notion of the Vaidya spacetime is expanded in the generalized version to encompass a wide range of scenarios, accommodating different forms of matter and radiation.   The spacetime is dynamic and undergoes evolution as matter compresses, with the metric describing the changing curvature \cite{husain,wang}.   Within the framework of the generalized Vaidya spacetime, employing comoving observers that satisfy the condition ``$dt = d\tau$" simplifies the mathematical representation of the spacetime.   It enables us to utilize a temporal reference that tracks the movement of matter as it undergoes gravitational collapse to become a black hole or emits radiation as a star.   Using a time parameter that evolves with the behavior of matter is a practical approach for investigating gravitational collapse or radiating stars. It improves the intuitiveness and physical significance of describing the collapse process.   When examining geodesic structures in the generalized Vaidya spacetime from the perspective of a comoving observer, our focus lies on the trajectories that objects or particles take when they deal with the changing spacetime caused by the reducing matter.   These geodesics illustrate the paths that things follow as they move through spacetime, which is influenced by changes in curvature caused by the dynamics of matter.   Comprehending these geodesic structures is essential for analyzing the dynamics of particles, photons, and observers in spacetime.  It facilitates forecasting the movement and interaction of objects inside the gravitational field generated by collapsing matter and is a crucial component in the analysis of the physics and astrophysical phenomena occurring in these spacetimes.\\

Using Eqs. (\ref{28}), (\ref{31}) in Eq. (\ref{23}), we have
\ben
\frac{dr}{dt}+E(t)=1-\frac{2M}{r} \label{32}
\een
On the other hand, using Eq. (\ref{31}) in Eq. (\ref{25}), we get
\ben
\frac{dr}{dt}+E(t)=2\mathcal{L}-\frac{L^{2}}{r^{2}}
\label{33}
\een
By substituting Eqs. (\ref{24}) and (\ref{30}) into Eq. (\ref{33}), we obtain the angular relation as follows:
\ben
r+L\Phi=2\mathcal{L}t+t_{0}E(t)
\label{34}
\een
taking integration constant to be zero.

For {\it non-radial} geodesic, using Eqs. (\ref{32}) and (\ref{33}) we have 
\ben
\Big(\frac{dr}{dt}+E(t) \Big)^{2}+\Big(D^{2}-2 \Big)\Big(\frac{dr}{dt}+E(t) \Big)+\Big(1-2D^{2}\La\Big)=0\nonumber\\
\label{35}
\een
where $D=\frac{2M}{L}$ and $L\neq 0$. Solving the above Eq. (\ref{35}), we obtain
\ben
\frac{dr}{dt}+E(t)=\frac{1}{2}\Big[(2-D^{2})\pm D\sqrt{D^{2}-4(1-2\La)}\Big]
\label{36}
\een
where we tate only positive solutions for our study.  By employing Eq. (\ref{30}) and performing integration on the aforementioned Eq. (\ref{36}), we obtain
\ben
r=t_{0}E(t)+\frac{t}{2}\Big[(2-D^{2})\pm D\sqrt{D^{2}-4(1-2\La)}\Big]+c\nonumber\\
\label{37}
\een
where $c$ is an integration constant.

\subsubsection{Time-like Geodesics for Case-I}
In order to analyze the structure of the time-like geodesics in the specified spacetime (\ref{13}) with the mass function (\ref{27}), we impose the condition $2\La\equiv \bar{G}_{\mu\nu}\dot{x}^{\mu}\dot{x}^{\nu}=-1$.  In this particular circumstance, Eq. (\ref{33}) is transformed as
\ben
\frac{dr}{dt}+E(t)=-1-\frac{L^{2}}{r^{2}}.
\label{38}
\een

First, we will analyze the radial geodesics with $L=0$, and then we will go on to the non-radial geodesics with $L\neq 0$. These geodesics are studied from the perspective of time-like geodesics of the generalized K-essence Vaidya metric (\ref{13}), considering the mass function (\ref{27}), inside a comoving system.

In order to track the {\it radial} geodesics ($\dot{\Phi}=0$), we consider the motion of a particle with no angular momentum ($L=0$) that starts its journey from a state of rest at a distance of $r=r_{a}$ and time $t=t_{a}$, so that the rate of change of its radial position with respect to time, $\frac{dr}{dt}$, is zero.   Thus, by referring to Eq. (\ref{32}), we obtain  
\ben
r_{a}=\frac{2M}{1-E(t_{a})}.
\label{39}
\een

Using Eq. (\ref{30}) in Eq. (\ref{38}), we get
\ben
r=-t+t_{0}E(t)+c_{1}
\label{40} 
\een
where $c_{1}$ is an integration constant. Using the aforementioned two Eqs. (\ref{39}) and (\ref{40}) in conjunction with the previously mentioned radial geodesics criteria for a particle, we get the expression for constant $c_{1}$ as
\ben
c_{1}=\frac{2M}{1-E(t_{a})}+t_{a}-t_{0}E(t_{a}) 
\label{41}
\een

Hence from Eq. (\ref{40}), we obtain
\ben
r=-t+t_{0}E(t)+\frac{2M}{1-E(t_{a})}+t_{a}-t_{0}E(t_{a}).
\label{42}
\een

Now, we have the capability to compute the duration it takes for a particle to reach the singularity ($r=0$) at a specific time $t=t_{s}$ along the radial geodesic in our given spacetime which is 
\ben
&&-t_{s}+t_{0}E(t_{s})+\frac{2M}{1-E(t_{a})}+t_{a}-t_{0}E(t_{a})=0\nonumber\\
&&\implies  t_{s}= t_{0}\ln{\Big[ \frac{1}{2W_{0}\Big(\frac{1}{2} e^{\frac{p}{t_{0}}}\Big)}\Big]}  \label{43}
\een
where $p=\big(-\frac{2M}{1-E(t_{a})}-t_{a}+t_{0}E(t_{a})\big)$ and $W_{0}(e^{\frac{p}{t_{0}}})$ is the Lambert W function \cite{Corless} provided that $e^{\frac{p}{t_{0}}}\geq 0$. Additionally, $W_{0}(q)$ is the solution of the equation $xe^{x}=q$ when $q$ is a non-negative real number. It is also noted that the Lambert $W_{0}$ function is a single-valued function. It is important to note that Eq. (\ref{35}) can be used to investigate radial geodesics. Because $D\to \infty$ when $L\to 0$ i.e., $\frac{1}{D}\to 0$ when $L\to 0$. So that the Eq. (\ref{35}) transformed to 
\ben
\frac{dr}{dt}+E(t)=2\mathcal{L}
\label{44}
\een
so that for radial time-like geodesics if we substitute $2\mathcal{L}=-1$ then it the exactly same with Eq. (\ref{38}) for $L=0$.\\

At the moment, we are tracking the {\it non-radial} scenario using the time-like geodesic framework. In this study, we use the Eqs. (\ref{35}) and (\ref{36}).\\

{\it Case-A:} First, look at $D= 2\sqrt{2}$ for the real root of the Eq. (\ref{36}), and then we use Eq. (\ref{30}) to find 
\ben
r-t_{0}E(t)+3t-c_{2}=0
\label{45}
\een
where $c_{2}$ is an integration constant, 
and from Eq. (\ref{34}) we get
\ben
\Phi=\frac{1}{L}\Big(2t-c_{2} \Big). \label{46}
\een

Let a particle starts its journey from $r=r_{a}$ at time $t=t_{a}$, then from (\ref{45}) we obtain
\ben
c_{2}= r_{a}-t_{0}E(t_{a})+3t_{a}
\label{47}
\een
By observing that $\frac{t_{a}}{t_{0}}\ge 0$ and $0\le E(t_{a}) <\frac{1}{2}$, we obtain the condition as long as all the values are finite is
\ben
  c_{2} \le r_{a}+3t_{a} < \frac{t_{0}}{2}+c_{2}.  
  \label{48}
\een

So that the Eqs. (\ref{45}) and (\ref{46}) transformed to
\ben
& r-t_{0}E(t)+3t-r_{a}+t_{0}E(t_{a})-3t_{a}=0 \label{49}\\
& \Phi=\frac{1}{L}\Big(2t-r_{a}+t_{0}E(t_{a})-3t_{a} \Big). \label{50}
\een

If we assume that the particle is moving towards the singularity $(r\rightarrow 0)$ when time $t\rightarrow t_{s}$, then according to Eq. (\ref{49}), we obtain
\ben
 t_{s}= t_{0}\ln{\Big[ \frac{1}{6W_{0}\Big(\frac{1}{6} e^{\frac{p_{1}}{3t_{0}}}\Big)}\Big]} 
\label{51}
\een
where $p_{1}=-r_{a}+t_{0}E(t_{a})-3t_{a}$ and $e^{\frac{p_{1}}{3t_{0}}}\ge 0$.

\begin{figure}[h]
\centering
\includegraphics[width=9cm]{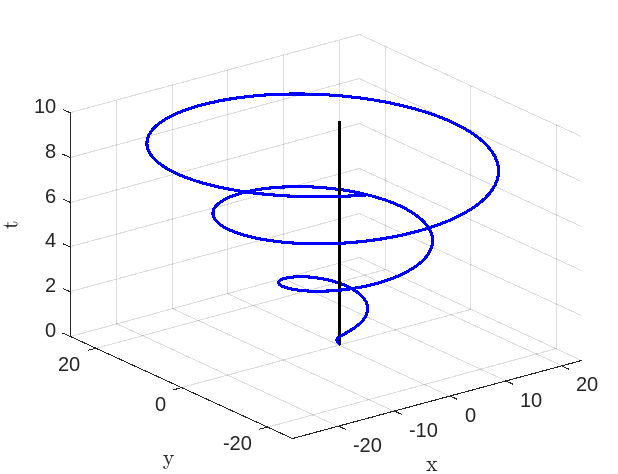}
\caption{Time-like geodesics for $D=2\sqrt{2}$, $L=1$, $t_{0}=0.1$, $c_{2}=0.1$}
\label{fig:1}
\end{figure}

\begin{figure}[h]
\centering
\includegraphics[width=9cm]{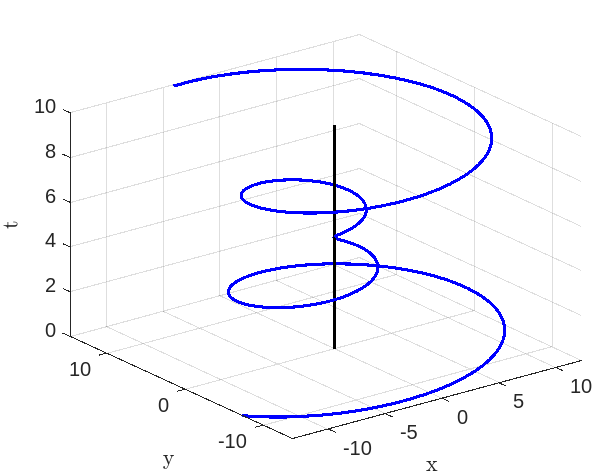}
\caption{Time-like geodesics for $D=2\sqrt{2}$, $L=1$, $t_{0}=0.1$, $c_{2}=15$}
\label{fig:2}
\end{figure}

Hence, Eqs. (\ref{45}) and (\ref{46}) demonstrate the possibility of tracing several non-radial time-like geodesics for varying values of $c_{2}$ under the condition $D=2\sqrt{2}$. As a result, a particle starts its journey from $r=r_{a}$ when $\Phi= \Phi_{a}=\frac{1}{L}\big(-r_{a}+t_{0}E(t_{a})-t_{a} \big)$, $t=t_{a}$ and will approach to singularity $r \rightarrow 0$ when $\Phi=\Phi_{s}=\frac{1}{L}\big(2t_{s}-r_{a}+t_{0}E(t_{a})-3t_{a} \big)$, $t=t_{s}= t_{s}= t_{0}\ln{\big[ \frac{1}{6W_{0}\big(\frac{1}{6} e^{\frac{p_{1}}{3t_{0}}}\big)}\big]} $ and it will leave the singularity when $t>t_{s}= t_{0}\ln{\big[ \frac{1}{6W_{0}\big(\frac{1}{6} e^{\frac{p_{1}}{3t_{0}}}\big)}\big]} $.
Given that $c_{2} \le r_{a}+3t_{a}< \frac{t_{0}}{2}+c_{2}$, it follows that the values of $r_{a}$ and $t_{a}$ are finite for all finite values of $c_{2}$ and $t_{0}$. Also, since $t_{s}= t_{0}\ln{\big[ \frac{1}{6W_{0}\big(\frac{1}{6} e^{\frac{p_{1}}{3t_{0}}}\big)}\big]}$, the value of $t_{s}$ is finite. By using Eqs. (\ref{45}) and (\ref{46}), we have plotted two non-radial time-like geodesics for $L=1$ and $t_{0}=0.1$, each corresponding to distinct values of $c_{2}=0.1$ and $c_{2}=15$. These geodesics are depicted in Figs.~(\ref{fig:1}) and~(\ref{fig:2}), respectively.   To plot the geodesics in these diagrams, we have converted the coordinate system from polar coordinates $(r,\Phi)$ to cartesian coordinates $(x,y)$. It should be noted that there exists central singularity for both the conventional generalized Vaidya spacetime \cite{husain,wang} and the generalized K-essence Vaidya spacetime \cite{gm3,gm4}, meaning that both $r\rightarrow 0$ and $t\rightarrow 0$. However, a singularity is typically defined as only $r\rightarrow 0$. In this case, we may see that the particle's track will allow a future observer to watch the particle reach $r\rightarrow 0$ at a specific time and escape the singularity. These phenomena are discussed in the conclusion section.

{\it Case-B:} To calculate the real solutions of Eq. ({\ref{35}}) for the purpose of analyzing non-radial time-like geodesics, we assume that $D> 2\sqrt{2}$. So that the Eq. (\ref{35}) can be written as
\ben
& \Big( \frac{dr}{dt}+E(t) -A\Big)\Big( \frac{dr}{dt}+E(t) -B\Big)=0\nonumber\\
\implies & \Big( r-t_{0}E(t) -At-c_{3}\Big)\Big( r-t_{0}E(t) -Bt-c_{3}\Big)=0\nonumber\\ 
\label{52}
\een
where 
\ben
& A=1-\frac{D^{2}}{2}+\frac{D}{2}\sqrt{D^{2}-8}~,\nonumber\\
&~B=1-\frac{D^{2}}{2}-\frac{D}{2}\sqrt{D^{2}-8}
\label{53}
\een
and $c_{3}$ is an integration constant.
        
The Eq. (\ref{52}) demonstrates that there are two non-radial time-like geodesics that may be followed for any finite value of $c_{3}$.

For
\ben
r-t_{0}E(t) -At-c_{3}=0 
\label{54}
\een 
then from Eq. (\ref{34}),
\ben
\Phi=\frac{1}{L}\Big[-t-At-c_{3}\Big] \label{55}
\een
and for 
\ben
r-t_{0}E(t) -Bt-c_{3}=0 
\label{56}
\een
we have from (\ref{34}),
\ben
\Phi=\frac{1}{L}\Big[-t-Bt-c_{3}\Big].\label{57}
\een
Let a particle starts its journey from $r=r_{a}$ in the path (\ref{54}) at time $t=t_{a}$, the from the equation (\ref{54})
\ben
c_{3}= r_{a}-t_{0}E(t_{a})-At_{a} 
\label{58}
\een
Clearly, since $\frac{t_{a}}{t_{0}}\ge 0$ and $0\le E(t_{a}) <\frac{1}{2}$ so that 
\ben
c_{3} \le r_{a}-At_{a} < \frac{t_{0}}{2}+c_{3}   \label{59}
\een
then the Eqs. (\ref{54}) and (\ref{55}) transformed to
\ben
& r-t_{0}E(t) -At-r_{a}+t_{0}E(t_{a})+At_{a}=0 \label{60}\\
& \Phi=\frac{1}{L}\Big[-t-At-r_{a}+t_{0}E(t_{a})+At_{a}\Big] \label{61}
\een
And if  we consider that the particle approaches to $r\rightarrow 0$ when $t\rightarrow t_{s}$ on the path (\ref{54}), then from (\ref{60}) we have
\ben
 t_{s}= t_{0}\ln{\Big[ \frac{1}{-2AW_{0}\Big(-\frac{1}{2A} e^{\frac{p_{2}}{At_{0}}}\Big)}\Big]}
\label{62}
\een
where $p_{2}=r_{a}-t_{0}E(t_{a})-At_{a}$ and $\frac{1}{A} e^{\frac{p_{2}}{At_{0}}}<0$.

Similarly, as before, let us consider a particle that begins its trajectory at $r=r_{a}$ along the path (\ref{56}) at time $t=t_{a}$ and eventually approaches singularity at $t=t_{s}$. By analysing equations (\ref{56}) and (\ref{57}), we may derive the following results:  
\ben
& r-t_{0}E(t) -Bt-r_{a}+t_{0}E(t_{a})+Bt_{a}=0 
\label{63}\\
& \Phi=\frac{1}{L}\Big[-t-Bt-r_{a}+t_{0}E(t_{a})+Bt_{a}\Big] \label{64}\\
&  c_{3}= r_{a}-t_{0}E(t_{a})-Bt_{a} \label{65}\\
&  c_{3} \le r_{a}-Bt_{a} < \frac{t_{0}}{2}+c_{3}   
\label{65}\\
& t_{s}= t_{0}\ln{\Big[ \frac{1}{-2BW_{0}\Big(-\frac{1}{2B} e^{\frac{p_{3}}{Bt_{0}}}\Big)}\Big]}
\label{67}
\een
where $p_{3}=r_{a}-t_{0}E(t_{a})-Bt_{a}$ and $\frac{1}{B} e^{\frac{p_{3}}{Bt_{0}}}\le 0$

\begin{figure}[h]
\centering
\includegraphics[width=9cm]{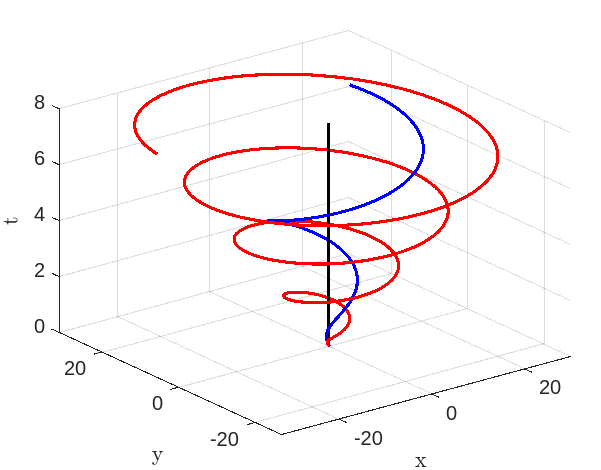}
\caption{Time-like geodesics for $D=2.9$, $L=1$, $t_{0}=0.1$, $c_{3}=0.1$: Blue Colour line for $r-t_{0}E(t) -At-c_{3}=0$ and Red colour line for $r-t_{0}E(t) -Bt-c_{3}=0$ }
\label{fig:3}
\end{figure}

\begin{figure}[h]
\centering
\includegraphics[width=9cm]{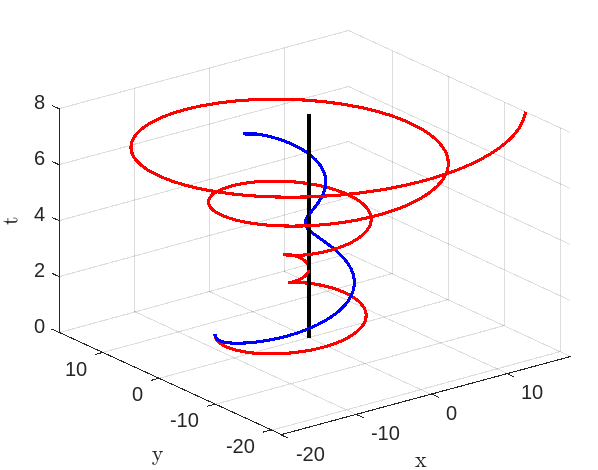}
\caption{Time-like geodesics for $D=2.9$, $L=1$, $t_{0}=0.1$, $c_{3}=10$: Blue Colour line for $r-t_{0}E(t) -At-c_{3}=0$ and Red colour line for $r-t_{0}E(t) -Bt-c_{3}=0$ }
\label{fig:4}
\end{figure}

Therefore, Eqs. (\ref{54}), (\ref{55}), (\ref{56}), and (\ref{57}) demonstrate the possibility of tracking several non-radial time-like geodesics for distinct values of $c$ given that $D>2\sqrt{2}$. Hence, for the path (\ref{54}), a particle starts its journey from $r=r_{a}$ when $\Phi= \Phi_{a}=\frac{1}{L}\big[-t_{a}-r_{a}+t_{0}E(t_{a})\big]$, $t=t_{a}$ and approaching to $r\rightarrow 0$ when  $\Phi=\Phi_{s}=\frac{1}{L}\big[-t_{s}-At_{s}-r_{a}+t_{0}E(t_{a})+At_{a}\big] $, $t=t_{s}= t_{0}\ln{\big[ \frac{1}{-2AW_{0}\big(-\frac{1}{2A} e^{\frac{p_{2}}{At_{0}}}\big)}\big]} $ and it will leave the singularity when $t>t_{s}=t_{0}\ln{\big[ \frac{1}{-2AW_{0}\big(-\frac{1}{2A} e^{\frac{p_{2}}{At_{0}}}\big)}\big]} $. Also, for the path (\ref{56}),  a particle starts its journey from $r=r_{a}$ when $\Phi= \Phi_{a}=\frac{1}{L}\big[-t_{a}-r_{a}+t_{0}E(t_{a})\big]$, $t=t_{a}$ and will arrive at singularity $r=0$ when  $\Phi=\Phi_{s}=\frac{1}{L}\big[-t_{s}-Bt_{s}-r_{a}+t_{0}E(t_{a})+Bt_{a}\big] $, $t=t_{s}=  t_{0}\ln{\big[ \frac{1}{-2BW_{0}\big(-\frac{1}{2B} e^{\frac{p_{3}}{Bt_{0}}}\big)}\big]}$ and it will leave the singularity when $t>t_{s}= t_{0}\ln{\big[ \frac{1}{-2BW_{0}\big(-\frac{1}{2B} e^{\frac{p_{3}}{Bt_{0}}}\big)}\big]}$.
Since $ c_{3} \le r_{a}-At_{a} < \frac{t_{0}}{2}+c_{3}   $ and $c_{3} \le r_{a}-Bt_{a} < \frac{t_{0}}{2}+c_{3}  $, then for any finite values of $c_{3}$, $t_{0}$, $A$ and $B$, the values of $r_{a}$ and $t_{a}$ are also finite. Using the Eqs. (\ref{54}), (\ref{55}) and (\ref{56}), (\ref{57}), for $D=2.9$, $L=1$ and $t_{0}=0.1$, we have traced two non-radial time-like for $c_{3}=0.1$ in Figure-(\ref{fig:3}) and also for $c_{3}=10$ in Figure-(\ref{fig:4}).


\subsubsection{Null Geodesics for Case-I}

In this subsection, we want to elucidate the null geodesic behavior of the generalized K-essence Vaidya spacetime, as described by Eq. (\ref{13}) and the mass function given by Eq. (\ref{27}). We assume that $2\La=0$. For this study, the Eqs. (\ref{33}) and (\ref{34}) reduces to  
\ben
& \frac{dr}{dt}+E(t)=-\frac{L^{2}}{r^{2}}
\label{68}\\
&r+L\Phi=t_{0}E(t).
\label{69}
\een
 
In order to track the {\it radial} geodesics, we assume that a particle with zero angular momentum ($L=0$) begins its motion from a state of rest at a distance of $r=r_{a}$ and time $t=t_{a}$, such that the rate of change of $r$ with respect to $t$ is zero ($\frac{dr}{dt}=0$).   Thus, based on Eq. \ref{32}, the value of $r_{a}$ may be determined as $\frac{2M}{1-E(t_{a})}$.   Furthermore, according to Eq. (\ref{68}),  
\ben
r=t_{0}E(t)+c_{4} \label{70}
\een
where $c_{4}$ is an integration constant.
At $t=t_{a}$, we obtain
\ben
c_{4}=\frac{2M}{1-E(t_{a})}-t_{0}E(t_{a})  \label{71}
\een
so that
\ben
r=t_{0}E(t)+\frac{2M}{1-E(t_{a})}-t_{0}E(t_{a}).
\label{72}
\een
Assuming that the particle reaches the singularity ($r=0$) at a specific time $t=t_{s}$, we have determined
\ben
t_{s}=t_{0}\ln{\Big[\frac{1-E(t_{a})}{2E(t_{a})\Big( 1-E(t_{a})\Big)-(4M/t_{0})}\Big]}.
\label{73}
\een

Now, we will examine the characteristics of {\it non-radial ($L\neq 0$)} geodesics inside the framework of null geodesics in the given spacetime (\ref{13}), which is governed by the particular mass function (\ref{27}).
The requirement for the existence of real roots of Eq. (\ref{35}) under the assumption that $2\La=0$ is that $D\geq 2$. Now we discuss the following two cases:\\

{\it Case-A:} When $D=2$, we have from Eq. (\ref{36})
\ben
r-t_{0}E(t)+t-c_{5}=0
\label{74}
\een
and from Eq. (\ref{34}) we get
\ben
\Phi=\frac{1}{L}\Big(t-c_{5} \Big) 
\label{75}
\een
Let a particle starts its journey from $r=r_{a}$ at time $t=t_{a}$, the from (\ref{74})
\ben
c_{5}= r_{a}-t_{0}E(t_{a})+t_{a} \label{76}
\een
It is evident that the inequality $\frac{t_{a}}{t_{0}}\ge 0$ is true, and we also have the condition $0\le E(t_{a}) <\frac{1}{2}$ so that
\ben
  c_{5} \le r_{a}+t_{a} < \frac{t_{0}}{2}+c_{5}   \label{77}
\een
then the Eqs. (\ref{74}) and (\ref{75}) transformed to
\ben
& r-t_{0}E(t)+t-r_{a}+t_{0}E(t_{a})-t_{a}=0 \label{78}\\
& \Phi=\frac{1}{L}\Big(t-r_{a}+t_{0}E(t_{a})-t_{a} \Big) \label{79}
\een
And if  we consider that the particle reaches the singularity $(r=0)$ at time $t=t_{s}$, then from (\ref{78}
) we have
\ben
 t_{s}= t_{0}\ln{\Big[ \frac{1}{2W_{0}\Big(\frac{1}{2} e^{\frac{p_{4}}{t_{0}}}\Big)}\Big]}
\label{80}
\een
where $p_{4}=-r_{a}+t_{0}E(t_{a})-t_{a}$ and $e^{\frac{p_{4}}{t_{0}}} \ge 0$.

\begin{figure}[h]
\centering
\includegraphics[width=9cm]{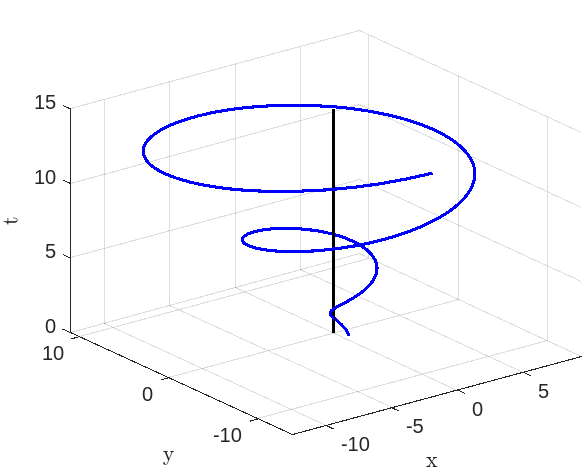}
\caption{Null geodesics for $D=2$, $L=1$, $t_{0}=0.1$, $c_{5}=1$}
\label{fig:5}
\end{figure}

\begin{figure}[h]
\centering
\includegraphics[width=9cm]{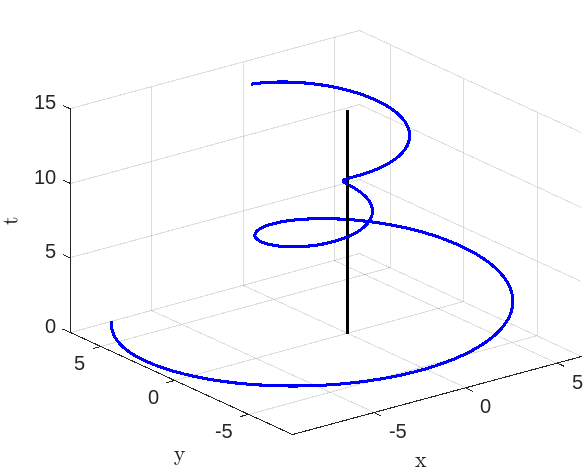}
\caption{Null geodesics for $D=2$, $L=1$, $t_{0}=0.1$, $c_{5}=10$}
\label{fig:6}
\end{figure}

Therefore, Eqs. (\ref{74}) and (\ref{75}) demonstrate that we may track several non-radial null geodesics for varying values of $c_{5}$ in the case of $D=2$. Hence, a particle starts its trajectory at a distance of $r=r_{a}$ when $\Phi= \Phi_{a}=\frac{1}{L}\big[-r_{a}+t_{0}E(t_{a})\big]$, $t=t_{a}$ and will arrive at singularity $r=0$ when  $\Phi=\Phi_{s}=\frac{1}{L}\big(t_{s}-r_{a}+t_{0}E(t_{a})-t_{a} \big)$, $t=t_{s}= t_{0}\ln{\big[ \frac{1}{2W_{0}\big(\frac{1}{2} e^{\frac{p_{4}}{t_{0}}}\big)}\big]} $ and it will leave the singularity when $t>t_{s}=t_{0}\ln{\big[ \frac{1}{2W_{0}\big(\frac{1}{2} e^{\frac{p_{4}}{t_{0}}}\big)}\big]} $. Given that $c_{5} \le r_{a}+t_{a}< \frac{t_{0}}{2}+c_{5}$, it follows that both $r_{a}$ and $t_{a}$ are finite for any finite values of $c_{5}$ and $t_{0}$. Also, since $t_{s}=t_{0}\ln{\big[ \frac{1}{2W_{0}\big(\frac{1}{2} e^{\frac{p_{4}}{t_{0}}}\big)}\big]}$, the value of $t_{s}$ is finite. By using Eqs. (\ref{74}) and (\ref{75}), we have plotted two non-radial time-like geodesics for $L=1$ and $t_{0}=0.1$, with two distinct values of $c_{5}$, namely $c_{5}=1$ and $c_{5}=10$. These geodesics are depicted in Figure-(\ref{fig:5}) and Figure-(\ref{fig:6}) correspondingly.\\
             
{\it Case-B:} When $D>2$, we have from Eq. (\ref{35})
\ben
\Big( r-t_{0}E(t) -At-c_{6}\Big)\Big( r-t_{0}E(t) -Bt-c_{6}\Big)=0,
\label{81}
\een
where 
\ben
&&A=1-\frac{D^{2}}{2}+\frac{D}{2}\sqrt{D^{2}-4},\nonumber\\&&B=1-\frac{D^{2}}{2}-\frac{D}{2}\sqrt{D^{2}-4}
\label{82}
\een
and $c_{6}$ is an integration constant. The Eq. (\ref{81}) demonstrates the existence of two non-radial null geodesics that may be followed for any finite value of $c_{6}$.   The Eq. (\ref{81}) is of the same form as (\ref{52}) in the context of time-like geodesics, but with different constants.   Therefore, based on this closeness, we may infer that equations (\ref{75}) and (\ref{81}) as
\ben
& r-t_{0}E(t) -At-c_{6}=0 \label{83}\\
& \Phi=\frac{1}{L}\Big(-At-c_{6} \Big).
\label{84}
\een  
Also,
\ben
& r-t_{0}E(t) -Bt-c_{6}=0 \label{85}\\
& \Phi=\frac{1}{L}\Big(-Bt-c_{6} \Big).
\label{86}
\een 
Let a particle starts its journey from $r=r_{a}$ in the path (\ref{83}) or (\ref{85}) at time $t=t_{a}$, then from the Eq. (\ref{83}), we have 
\ben
 & c_{6}= r_{a}-t_{0}E(t_{a})-At_{a} \label{87}\\
\text{and}~&  c_{6} \le r_{a}-At_{a} < \frac{t_{0}}{2}+c_{6}   \label{88}
\een
also 
 then from the equation (\ref{85}), we have 
\ben
& c_{6}= r_{a}-t_{0}E(t_{a})-Bt_{a} \label{89}\\
\text{and}~&  c_{6} \le r_{a}-Bt_{a} < \frac{t_{0}}{2}+c_{6}  \label{90}
\een    
Therefore we have
\ben
& r-t_{0}E(t) -At-r_{a}+t_{0}E(t_{a})+At_{a}=0 \label{91}\\
& \Phi=\frac{1}{L}\Big[-At-r_{a}+t_{0}E(t_{a})+At_{a}\Big] \label{92}
\een
and also 
\ben
& r-t_{0}E(t) -Bt-r_{a}+t_{0}E(t_{a})+Bt_{a}=0 \label{93}\\
& \Phi=\frac{1}{L}\Big[-Bt-r_{a}+t_{0}E(t_{a})+Bt_{a}\Big] \label{94}
\een
If we consider that the particle approaches the singularity $(r\rightarrow 0)$ at time $t=t_{s}$ along the path (\ref{83}) or (\ref{85}), then we may deduce from equation (\ref{91})
\ben
 t_{s}= t_{0}\ln{\Big[ \frac{1}{-2AW_{0}\Big(-\frac{1}{2A} e^{\frac{p_{5}}{At_{0}}}\Big)}\Big]}
\label{95} 
\een
and from (\ref{93}) 
 \ben
 t_{s}= t_{0}\ln{\Big[ \frac{1}{-2BW_{0}\Big(-\frac{1}{2B} e^{\frac{p_{6}}{Bt_{0}}}\Big)}\Big]}
\label{96}
 \een
where $p_{5}=r_{a}-t_{0}E(t_{a})-At_{a}$, $p_{6}=r_{a}-t_{0}E(t_{a})-Bt_{a}$, $\frac{1}{A} e^{\frac{p_{5}}{At_{0}}}\le 0$ and $\frac{1}{B} e^{\frac{p_{6}}{Bt_{0}}}\le 0$.

\begin{figure}[h]
\centering
\includegraphics[width=9cm]{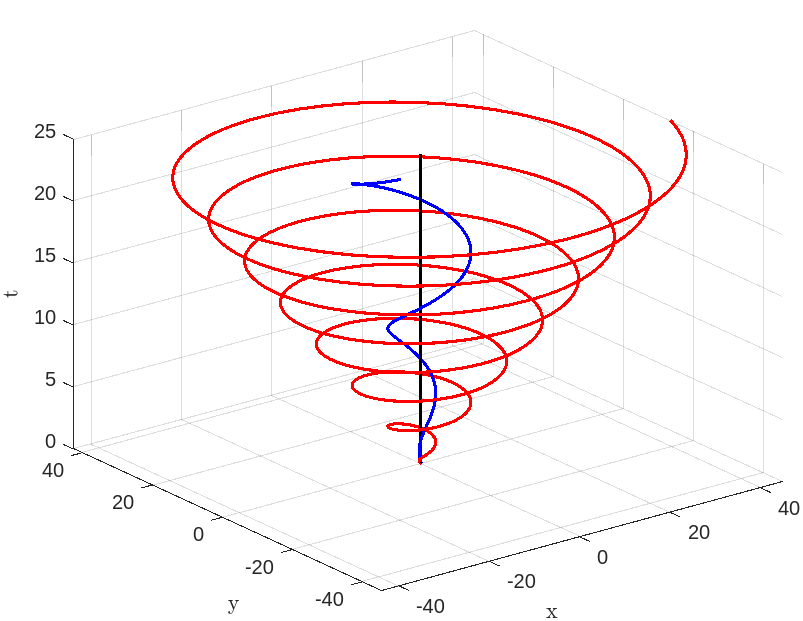}
\caption{Null geodesics for $D=2.1$, $L=1$, $t_{0}=0.1$, $c_{6}=0.1$: Blue Colour line for $r-t_{0}E(t) -At-c_{6}=0$ and Red colour line for $r-t_{0}E(t) -Bt-c_{6}=0$ }
\label{fig:7}
\end{figure}

\begin{figure}[h]
\centering
\includegraphics[width=9cm]{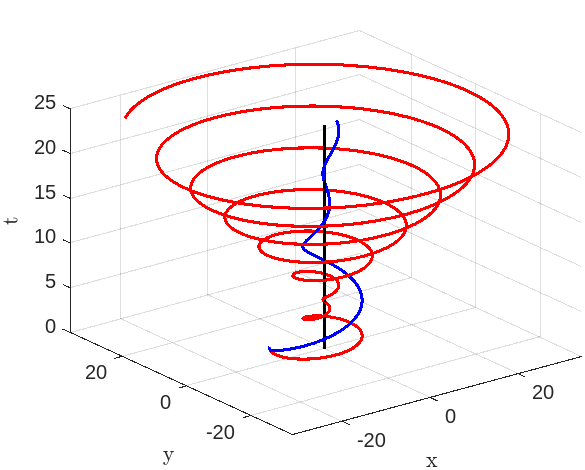}
\caption{Null geodesics for $D=2.1$, $L=1$, $t_{0}=0.1$, $c_{6}=10$: Blue Colour line for $r-t_{0}E(t) -At-c_{6}=0$ and Red colour line for $r-t_{0}E(t) -Bt-c_{6}=0$ }
\label{fig:8}
\end{figure}

As the outcome of the Eqs. (\ref{83}), (\ref{84}), and (\ref{85}), (\ref{86}), we may trace distinct non-radial null geodesics for different values of $c_6$ when $D>2$. As a result, given the path (\ref{83}), a particle begins its journey  from $r=r_{a}$ when $\Phi= \Phi_{a}=\frac{1}{L}\big[-r_{a}+t_{0}E(t_{a})\big]$, $t=t_{a}$ and will approaches singularity $r\rightarrow 0$ when  $\Phi=\Phi_{s}=\frac{1}{L}\big[-At_{s}-r_{a}+t_{0}E(t_{a})+At_{a}\big] $, $t=t_{s}= t_{0}\ln{\big[ \frac{1}{-2AW_{0}\big(-\frac{1}{2A} e^{\frac{p_{5}}{At_{0}}}\big)}\big]}$ and it will leave from the singularity when $t>t_{s}=t_{0}\ln{\big[ \frac{1}{-2AW_{0}\big(-\frac{1}{2A} e^{\frac{p_{5}}{At_{0}}}\big)}\big]} $.\\
Also, for the path (\ref{85}),  a particle starts its journey from $r=r_{a}$ when $\Phi= \Phi_{a}=\frac{1}{L}\big[-r_{a}+t_{0}E(t_{a})\big]$, $t=t_{a}$ and will approaches singularity $r\rightarrow 0$ when  $\Phi=\Phi_{s}=\frac{1}{L}\big[-Bt_{s}-r_{a}+t_{0}E(t_{a})+Bt_{a}\big] $, $t=t_{s}= t_{0}\ln{\big[ \frac{1}{-2BW_{0}\big(-\frac{1}{2B} e^{\frac{p_{6}}{Bt_{0}}}\big)}\big]} $ and it will leave the singularity when $t>t_{s}=t_{0}\ln{\big[ \frac{1}{-2BW_{0}\big(-\frac{1}{2B} e^{\frac{p_{6}}{Bt_{0}}}\big)}\big]}$.\\
 Since $ c_{6} \le r_{a}-At_{a} < \frac{t_{0}}{2}+c_{6}   $ and $c_{6} \le r_{a}-Bt_{a} < \frac{t_{0}}{2}+c_{6}  $, then for any finite values of $c_{6}$, $t_{0}$, $A$ and $B$, the values of $r_{a}$ and $t_{a}$ are also finite. 
 By substituting the values $D=2.1$, $L=1$, and $t_{0}=0.1$ into equations (\ref{83}), (\ref{84}), (\ref{85}), and (\ref{86}), we have obtained two non-radial nulls for $c_{6}=0.1$ in Figure-(\ref{fig:7}) and also for $c_{6}=10$ in Figure-(\ref{fig:8}).

\subsection{Case-II: $\mathcal{M}(t,r)=\mu t+\frac{r}{2}\phi_{t}^{2}$}
In this part, we consider the generalized K-essence Vaidya mass function (\ref{14}) with the assumption (\ref{26}) as 
\ben
\mathcal{M}(t,r)=\mu t+\frac{r}{2}e^{-\frac{t}{t_{0}}}
\label{97}
\een
where we take the usual generalized Vaidya mass function $m(t,r)\equiv m(t)=\mu t$ i.e., is a linear function of $t$ \cite{vert1,Solanki,Blau} and $\mu$ is a positive constant.

From the above mass function (\ref{97}), we have 
\ben
\frac{\mathcal{\bar{M}}_{t}}{r}=-\frac{e^{-\frac{t}{t_{0}}}}{2t_{0}}
\label{98}
\een
where we redefine the first derivative of the given mass function with respect to time as $\mathcal{\bar{M}}_{t}\equiv\mathcal{M}_{t}-\mu$.

Following the same procedure as for Case-I, and using Eqs. (\ref{20}-\ref{25}) and (\ref{28}), and again assuming the comoving plane ($d\tau\equiv dt$), we obtain 
\ben
&  E(t)=\frac{1}{2} e^{-\frac{t}{t_{0}}} \label{99}\\
\text{and}~~&\mathcal{M}(t.r)=\mu t+E(t)r. \label{100}
\een

Again, continuing the similar procedure as {\it Case-I}, we found the following relations:
\ben
& \frac{dr}{dt}+E(t)=2\mathcal{L}-\frac{L^{2}}{r^{2}} \label{101}\\
& \frac{dr}{dt}+E(t)=1-\frac{2\mu t}{r} \label{102}\\
& r+L\Phi=2\mathcal{L}t+t_{0}E(t)\label{103}
\een

\subsubsection{Time-like Geodesics for Case-II}
In this portion, we analyze the behavior of time-like geodesics in the generalized K-essence Vaidya spacetime, as described by Eq. (\ref{13}) with the mass function given by Eq. (\ref{97}). For the sake of this investigation, we assume $2\La=-1$.\\

Following from the previous analysis for {\it Case-I}, we will now focus on tracking the radial geodesics. Specifically, we will examine a particle with zero angular momentum ($L=0$) that begins its motion from a state of rest at a distance $r=r_{a}$ and time $t=t_{a}$, 
\ben
r_{a}=\frac{2\mu t_{a}}{1-E(t_{a})}   
\label{104}
\een
and using Eq. (\ref{38}), we obtain 
\ben
& r=-t+t_{0}E(t)+c_{7}\label{105}\\
& c_{7}=\frac{2\mu t_{a}}{1-E(t_{a})} +t_{a}-t_{0}E(t_{a})\label{106}\\
& r=-t+t_{0}E(t)+\frac{2\mu t_{a}}{1-E(t_{a})} +t_{a}-t_{0}E(t_{a})\label{107}
\een
Similarly, if the particle approaches the singularity ($r\rightarrow 0$) at time $t=t_{s}$, the same type of consideration applies, we get
\ben
t_{s}=  t_{0}\ln{\Big[ \frac{1}{2W_{0}\Big(\frac{1}{2} e^{\frac{p_{7}}{t_{0}}}\Big)}\Big]}  \label{108}
\een
provided $e^{\frac{p_{7}}{t_{0}}}\ge 0$ where $p_{7}=-\frac{2\mu t_{a}}{1-E(t_{a})} -t_{a}+t_{0}E(t_{a})$.

When studying non-radial time-like geodesics, we obtain the same expression as Eq. (\ref{35}), but with a different value for the parameter $D$ ($D=\frac{2\mu t}{L}$).   In order to trace the geodesics, it is necessary to satisfy the given requirement
\ben
D\ge 2\sqrt{2} \implies t\ge \frac{\sqrt{2}L}{\mu}.
\label{109}
\een

In this non-radial study, the scenario where $D=2\sqrt{2}$ cannot be taken into consideration. This is because if $D=2\sqrt{2}$, then $t=\frac{\sqrt{2}L}{\mu}$, resulting in a constant value for time. As a result, the spacetime is not like the generalized K-essence Vaidya type. So, in this non-radial time-like geodesic investigation, we only look into $D> 2\sqrt{2}$, ensuring that the spacetime is time-dependent, which is an essential characteristic of both the normal Vaidya spacetime and the generalized K-essence Vaidya spacetime. As before, we solve the Eq. (\ref{35}) for $D> 2\sqrt{2}$ where $D=\frac{2\mu t}{L}$, we obtain
\ben
\Big( r-t_{0}E(t) -\bar{A}(t)-c_{8}\Big)\Big( r-t_{0}E(t) -\bar{B}(t)-c_{8}\Big)=0 \nonumber\\
\label{110}
\een
where 
\ben
& \bar{A}(t) =t-\frac{2\mu^{2}t^{3}}{3L^{2}}+\frac{2\mu^{2}}{3L^{2}}\Big(t^{2}-\frac{2L^{2}}{\mu^{2}} \Big)^{\frac{3}{2}}\label{111}\\
& \bar{B}(t) =t-\frac{2\mu^{2}t^{3}}{3L^{2}}-\frac{2\mu^{2}}{3L^{2}}\Big(t^{2}-\frac{2L^{2}}{\mu^{2}} \Big)^{\frac{3}{2}}.
\label{112}
\een

Clearly, Eq. (\ref{110}) shows that there are two non-radial time-like geodesics that can be traced for any finite value of $c_{8}$. Therefore, we have two non-radial time-like geodesics, either,
\ben
& r-t_{0}E(t) -\bar{A}(t)-c_{8}=0 \label{113}\\
& \Phi=\frac{1}{L}\Big[-t-\bar{A}(t)-c_{8}\Big], \label{114}
\een
or, 
\ben
& r-t_{0}E(t) -\bar{B}(t)-c_{8}=0 \label{115}\\
& \Phi=\frac{1}{L}\Big[-t-\bar{B}(t)-c_{8}\Big] \label{116}
\een

Let a particle starts its journey from $r=r_{a}$ in the path (\ref{113}) at time $t=t_{a} > \frac{\sqrt{2}L}{\mu}$, we have
\ben
c_{8}= r_{a}-t_{0}E(t_{a})-\bar{A}(t_{a}) \label{117}
\een
Since $\frac{t_{a}}{t_{0}}\ge 0$ and $0\le E(t_{a}) <\frac{1}{2}$
\ben
c_{8} \le r_{a}-\bar{A}(t_{a}) < \frac{t_{0}}{2}+c_{8}   \label{118}
\een
then the Eqs. (\ref{113}) and (\ref{114}) transformed to
\begin{align}
& r-t_{0}E(t) -\bar{A}(t)-r_{a}+t_{0}E(t_{a})+\bar{A}(t_{a})=0 \label{119}\\
& \Phi=\frac{1}{L}\Big[-t-\bar{A}(t)-r_{a}+t_{0}E(t_{a})+\bar{A}(t_{a})\Big] \label{120}
\end{align}
then at $t=t_{a}$ 
\begin{align}
\Phi=\Phi_{a}=\frac{1}{L}\Big[-t_{a}-r_{a}+t_{0}E(t_{a})\Big]. \label{121}
\end{align}
If we assume that the particle reaches the singularity $(r\rightarrow 0)$ at time $t=t_{s}$ along the path (\ref{113}), then we can deduce from Eq. (\ref{119}) the following:
\begin{align}
&  t_{0}E(t_{s}) +\bar{A}(t_{s})+c_{8}=0\nonumber\\
 &  E(t_{s}) -\frac{2\mu^{2}t^{3}_{s}}{3t_{0}L^{2}}+\frac{c_{8}}{t_{0}}=-\frac{t_{s}}{t_{0}}-\frac{2\mu^{2}}{3t_{0}L^{2}}\Big(t^{2}_{s}-\frac{2L^{2}}{\mu^{2}} \Big)^{\frac{3}{2}} 
\label{122}
\end{align}

\begin{figure}[h]
\centering            \includegraphics[width=9cm]{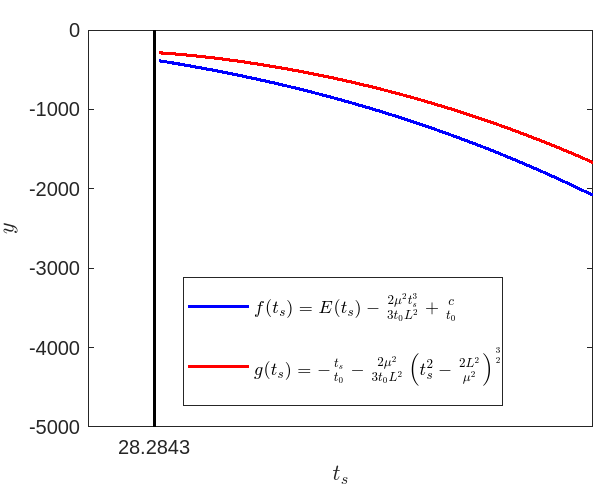}
\caption{$\mu=0.05$, $L=1$, $t_{0}=0.1$ and $c_{8}=0.1$: Blue Colour line for $f(t_{s})$ and Red colour line for $g(t_{s})$ }
\label{fig:9}
\end{figure}

\begin{figure}[h]
\centering            \includegraphics[width=9cm]{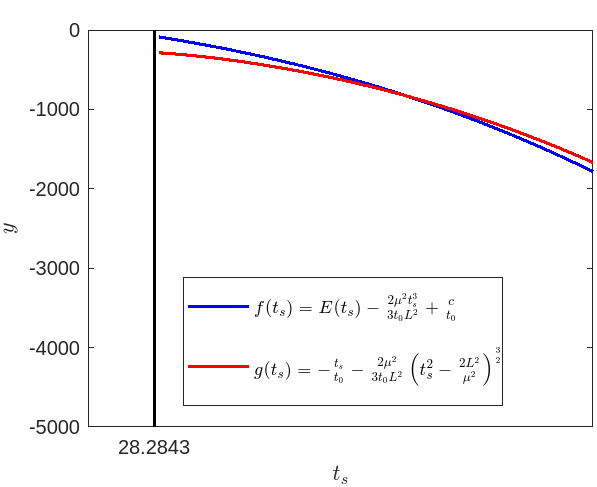}
\caption{$\mu=0.05$, $L=1$, $t_{0}=0.1$ and $c_{8}=30$: Blue Colour line for $f(t_{s})$ and Red colour line for $g(t_{s})$ }
\label{fig:10}
\end{figure}

\begin{figure}[h]
\centering            \includegraphics[width=9cm]{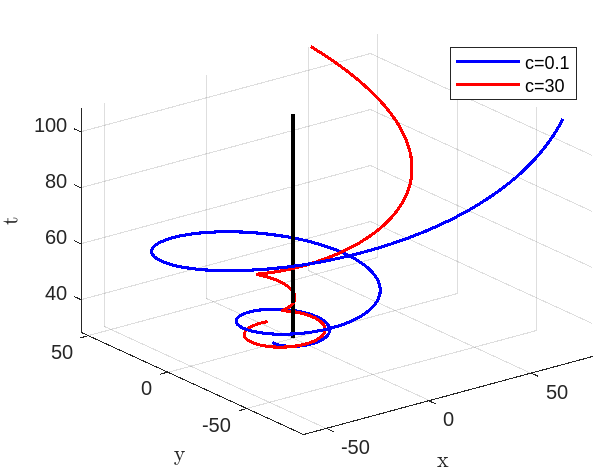}
\caption{Time-like geodesics for $r-t_{0}E(t) -\bar{A}-c_{8}=0$: When $\mu=0.05$, $L=1$, $t_{0}=0.1$: Blue Colour line for $c_{8}=0.1$ and Red colour line for $c_{8}=30$ }
\label{fig:11}
\end{figure}

In this scenario, the precise expression for $t_{s}$ cannot be determined due to the presence of a distinct mass function (\ref{97}). However, an expression for the transcendental equation (\ref{112}) is available, allowing for numerical and graphical analysis to get the finite time for any given values of $\mu$ and $L$. Now let $f(t_{s})=E(t_{s}) -\frac{2\mu^{2}t^{3}_{s}}{3t_{0}L^{2}}+\frac{c}{t_{0}}$ and $g(t_{s})=-\frac{t_{s}}{t_{0}}-\frac{2\mu^{2}}{3t_{0}L^{2}}\big(t^{2}_{s}-\frac{2L^{2}}{\mu^{2}} \big)^{\frac{3}{2}} $. Then the only possibility for a finite value $t_{s}$ if and only if $f(t_{s}) = g(t_{s})$. \\
To start with let us consider $\mu=0.05$ and $L=1$ then $t_{a}>28.2843$. Let $t_{a}=28.3$ Then from (\ref{111}), $\bar{A}(t_{a})\approx-9.4739$. Now if we consider $t_{0}=0.1$ then for $c_{8}=0.1$, $r_{a}\approx-9.3739$  $\Phi_{a}\approx -18.9261$ and also for $c_{8}=30$, $r_{a}\approx 20.5261$ and $\Phi_{a}\approx -48.8261$ by using (\ref{117}) and (\ref{118}). Figure-(\ref{fig:9}) shows that the {\it for $c_{8}=0.1$,} the Eq. (\ref{122}) has no solution for $t_{s}$ since the  curves $y=f(t_{s})$ and $y=g(t_{s})$ do not intersect each other. So for $c_{8}=0.1$, the particle which starts from $r=r_{a}$ at time $t=t_{a}>28.2843$ but it will never reach the singularity as shown in Figure-(\ref{fig:11}) (blue line) using the Eqs. (\ref{113}) and (\ref{114}). However, for $c_{8}=30$, from Figure-(\ref{fig:10}) shows that it has a solution for $t_{s}$ since the the curves $y=f(t_{s})$ and $y=g(t_{s})$ intersect each other.  For $c_8=30$, the particle initially located at $r=r_a$ at time $t=t_a>28.2843$ will eventually reach the singularity ($r=0$) at time $t=t_s$, as depicted by the red line in Figure-(\ref{fig:11}). The particle will then depart from the singularity ($r=0$), which is seen graphically by Eqs. (\ref{113}) and (\ref{114}).

Again, if we consider a particle starts its journey from $r=r_{b}$ in the path (\ref{115}) at time $t=t_{b} > \frac{\sqrt{2}L}{\mu}$, we have
\begin{align}
c_{8}= r_{b}-t_{0}E(t_{b})-\bar{B}(t_{b}) \label{123}
\end{align}
In this path, the condition is  $\frac{t_{b}}{t_{0}}\ge 0$ and $0\le E(t_{b}) <\frac{1}{2}$
\begin{align}
  c_{8} \le r_{b}-\bar{B}(t_{b}) < \frac{t_{0}}{2}+c_{8}   \label{124}
\end{align}
then the Eqs. (\ref{115}) and (\ref{116}) transformed to
\begin{align}
& r-t_{0}E(t) -\bar{B}(t)-r_{b}+t_{0}E(t_{b})+\bar{B}(t_{b})=0 \label{125}\\
& \Phi=\frac{1}{L}\Big[-t-\bar{B}(t)-r_{b}+t_{0}E(t_{b})+\bar{B}(t_{b})\Big] \label{126}
\end{align}
then at $t=t_{b}$ 
\begin{align}
\Phi=\Phi_{b}=\frac{1}{L}\Big[-t_{b}-r_{b}+t_{0}E(t_{b})\Big]. \label{127}
\end{align}

Once more, assuming it is possible, let us contemplate the scenario in which the particle moves towards the singularity $(r\rightarrow 0)$ at time $t=t_{s}$ along the trajectory (\ref{115}). Consequently, based on equation (\ref{125}), we may establish the following relation:
\begin{align}
E(t_{s}) -\frac{2\mu^{2}t^{3}_{s}}{3t_{0}L^{2}}+\frac{c_{8}}{t_{0}}=-\frac{t_{s}}{t_{0}}+\frac{2\mu^{2}}{3t_{0}L^{2}}\Big(t^{2}_{s}-\frac{2L^{2}}{\mu^{2}} \Big)^{\frac{3}{2}} 
\label{128}
\end{align}

\begin{figure}[h]
\centering            \includegraphics[width=9cm]{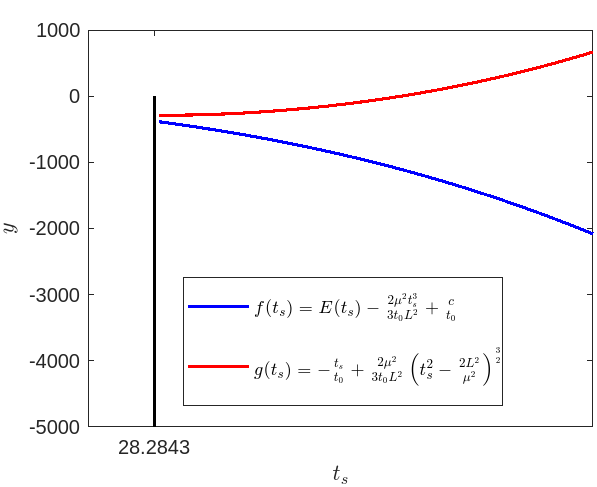}
\caption{$\mu=0.05$, $L=1$, $t_{0}=0.1$ and $c_{8}=0.1$: Blue Colour line for $f(t_{s})$ and Red colour line for $g(t_{s})$ }
\label{fig:12}
\end{figure}

\begin{figure}[h]
\centering            \includegraphics[width=9cm]{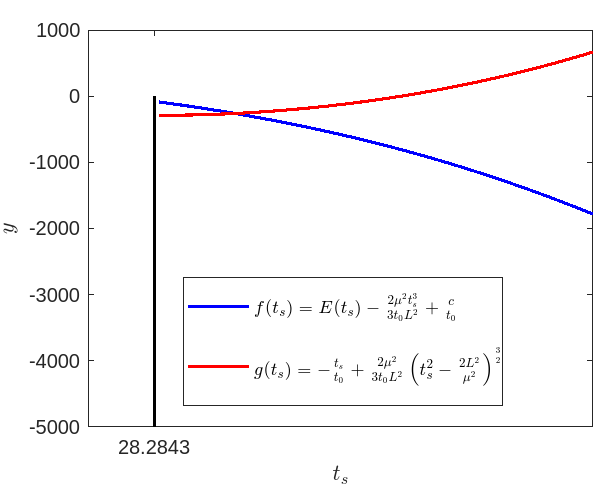}
\caption{$\mu=0.05$, $L=1$, $t_{0}=0.1$ and $c_{8}=30$: Blue Colour line for $f(t_{s})$ and Red colour line for $g(t_{s})$ }
\label{fig:13}
\end{figure}

\begin{figure}[h]
\centering            \includegraphics[width=9cm]{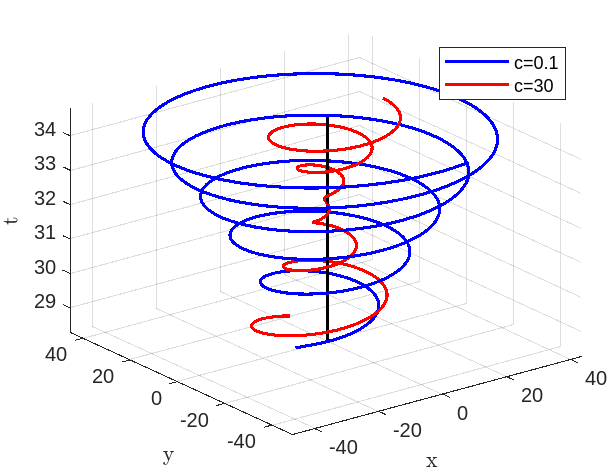}
\caption{Time-like geodesics for $r-t_{0}E(t) -\bar{B}-c_{8}=0$: When $\mu=0.05$, $L=1$, $t_{0}=0.1$: Blue Colour line for $c_{8}=0.1$ and Red colour line for $c_{8}=30$ }
\label{fig:14}
\end{figure}

Once again, this equation is transcendental, meaning it can be analyzed numerically and in pictures for any finite values of $\mu$ and $L$. Now let $f(t_{s})=E(t_{s}) -\frac{2\mu^{2}t^{3}_{s}}{3t_{0}L^{2}}+\frac{c_{8}}{t_{0}}$ and $g(t_{s})=-\frac{t_{s}}{t_{0}}+\frac{2\mu^{2}}{3t_{0}L^{2}}\big(t^{2}_{s}-\frac{2L^{2}}{\mu^{2}} \big)^{\frac{3}{2}} $. Then the only possibility for a finite value $t_{s}$ if and only if $f(t_{s}) = g(t_{s})$. \\
To start with, let us consider $\mu=0.05$ and $L=1$ then $t_{b}>28.2843$. Let $t_{b}=28.3$ Then, from (\ref{112}), $\bar{B}(t_{b})\approx-9.4767$. Now if we consider $t_{0}=0.1$ then for $c_{8}=0.1$, $r_{b}\approx -9.3767$  $\Phi_{b}\approx -18.9233$ and also for $c_{8}=30$, $r_{b}\approx 20.5233$ and $\Phi_{b}\approx -48.8233$ by using (\ref{123}) and (\ref{127}). Figure-(\ref{fig:12}) demonstrates that when $c_{8}=0.1$, the Eq. (\ref{128}) does not have a solution for $t_{s}$ because the curves $y=f(t_{s})$ and $y=g(t_{s})$ do not cross.  So for $c_{8}=0.1$, the particle which starts from $r=r_{b}$ at time $t=t_{b}>28.2843$ but it will never reach the singularity as shown in Figure-(\ref{fig:14}) (blue line) using the Eqs. (\ref{115}) and (\ref{116}). However, Figure-(\ref{fig:13}) shows that when $c_{8}=30$, Eq. (\ref{128}) has a solution for $t_{s}$ since the curves $y=f(t_{s})$ and $y=g(t_{s})$ intersect.   For a value of $c_8$ equal to 30, the particle begins at $r=r_b$ at a time $t=t_b$ greater than 28.2843. It will eventually reach the singularity at $r=0$ at time $t=t_s$, as depicted by the red line in Figure-(\ref{fig:14}). The particle will then depart from the location $r=0$, which can be determined from Figure-(\ref{fig:14}) using Eqs. (\ref{115}) and (\ref{116}).

\subsubsection{Null Geodesics for Case-II}

Within this part, we have studied the characteristics of both radial ($L=0$) and non-radial ($L\neq 0$) null geodesics in the generalized K-essence Vaidya spacetime (\ref{13}) with the specific mass function (\ref{97}).

By tracing the radial null geodesic using the same reasoning as previously, we obtain the identical value of $r_{a}$ as stated in Eq.(\ref{104}). Continuing in a similar manner as before, we have:
\begin{align}
r=t_{0}E(t)+\frac{2\mu t_{a}}{1-E(t_{a})}-t_{0}E(t_{a})
\label{129}
\end{align}
Again, we able to calculate the time $t=t_{s}$ for a particle approaches to singularity $r\rightarrow 0$, we obtain
\begin{align}
t_{s}= t_{0} \ln{\Big[ \frac{1- E(t_{a})}{2E(t_{a})[1- E(t_{a})]-\frac{4\mu t_{a}}{t_{0}}}  \Big]}.\label{130}
\end{align}

To trace the characteristics of non-radial ($L\neq 0$) null geodesics for the given spacetime (\ref{13}) through the Eq. (\ref{35}) we consider $D> 2 \implies  t\ge \frac{L}{\mu}$ ($D=\frac{2\mu t}{L}$). Similarly, we exclude the situation when $D=2$ due to the aforementioned rationale for non-radial time-like geodesics. Therefore, when $D> 2$, the Eq. (\ref{35}) can be written as
\begin{align}
 \Big( r-t_{0}E(t) -\bar{A}(t)-c_{9}\Big)\Big( r-t_{0}E(t) -\bar{B}(t)-c_{9}\Big)=0 \label{131}
\end{align}
where 
\begin{align}
& \bar{A}(t) =t-\frac{2\mu^{2}t^{3}}{3L^{2}}+\frac{2\mu^{2}}{3L^{2}}\Big(t^{2}-\frac{L^{2}}{\mu^{2}} \Big)^{\frac{3}{2}}\nonumber\\
& \implies \bar{A}(t) = \frac{2\mu^{2}}{3L^{2}}\mathcal{O}\Big(\frac{1}{t}\Big)\label{132}\\
~\text{and}~ 
& \bar{B}(t) =t-\frac{2\mu^{2}t^{3}}{3L^{2}}-\frac{2\mu^{2}}{3L^{2}}\Big(t^{2}-\frac{L^{2}}{\mu^{2}} \Big)^{\frac{3}{2}}\nonumber\\
& \implies \bar{B}(t) = 2t-\frac{4\mu^{2}t^{3}}{3L^{2}}-\frac{2\mu^{2}}{3L^{2}}\mathcal{O}\Big(\frac{1}{t}\Big)
\label{133}
\end{align}
The equation (\ref{131}) indicates the existence of two non-radial null geodesics that may be followed for any finite value of $c_{9}$. Proceeding as before, we have 
 \begin{align}
& r= t_{0}E(t)+\frac{2\mu^{2}}{3L^{2}}\mathcal{O}\Big(\frac{1}{t}\Big)+c_{9}\label{134}\\ 
\text{with}~~ &  \Phi= \frac{1}{L}\Big[-\frac{2\mu^{2}}{3L^{2}}\mathcal{O}\Big(\frac{1}{t}\Big)-c_{9}\Big]\label{135}\\
\text{and}~~ & r= t_{0}E(t)+2t-\frac{4\mu^{2}t^{3}}{3L^{2}}-\frac{2\mu^{2}}{3L^{2}}\mathcal{O}\Big(\frac{1}{t}\Big)+c_{9}\label{136}\\
\text{with}~~& \Phi=\frac{1}{L}\Big[-2t+\frac{4\mu^{2}t^{3}}{3L^{2}}+\frac{2\mu^{2}}{3L^{2}}\mathcal{O}\Big(\frac{1}{t}\Big)-c\Big]\label{137}
\end{align}

Again, we consider a particle starts its journey from $r=r_{a}$ in the path (\ref{134}) at time $t=t_{a} > \frac{L}{\mu}$, we get
\begin{align}
c_{9}= r_{a}-t_{0}E(t_{a})-\bar{A}(t_{a}) \label{138}
\end{align}
For this scenario, the condition imposed on the particle trajectory is as follows: 
\begin{align}
  c_{9} \le r_{a}-\bar{A}(t_{a}) < \frac{t_{0}}{2}+c_{9}   \label{139}
\end{align}
since $\frac{t_{a}}{t_{0}}\ge 0$ and $0\le E(t_{a}) <\frac{1}{2}$.
Hence, the Eqs. (\ref{134}) and (\ref{135}) becomes
\begin{align}
& r-t_{0}E(t) -\bar{A}(t)-r_{a}+t_{0}E(t_{a})+\bar{A}(t_{a})=0 \label{140}\\
& \Phi=\frac{1}{L}\Big[-\bar{A}(t)-r_{a}+t_{0}E(t_{a})+\bar{A}(t_{a})\Big] \label{141}
\end{align}
then at $t=t_{a}$ 
\begin{align}
    \Phi=\Phi_{a}=\frac{1}{L}\Big[-r_{a}+t_{0}E(t_{a})\Big] \label{142}
\end{align}
If it is achievable, let us assume that the particle travels along the path (\ref{134}) and reaches the singularity $(r=0)$ at time $t=t_{s}$. Then, from (\ref{140}) we get
\begin{align}
E(t_{s}) -\frac{2\mu^{2}t^{3}_{s}}{3t_{0}L^{2}}+\frac{c}{t_{0}}=-\frac{t_{s}}{t_{0}}-\frac{2\mu^{2}}{3t_{0}L^{2}}\Big(t^{2}_{s}-\frac{L^{2}}{\mu^{2}} \Big)^{\frac{3}{2}} 
\label{143}
\end{align}
 
\begin{figure}[h]
\centering            \includegraphics[width=8cm]{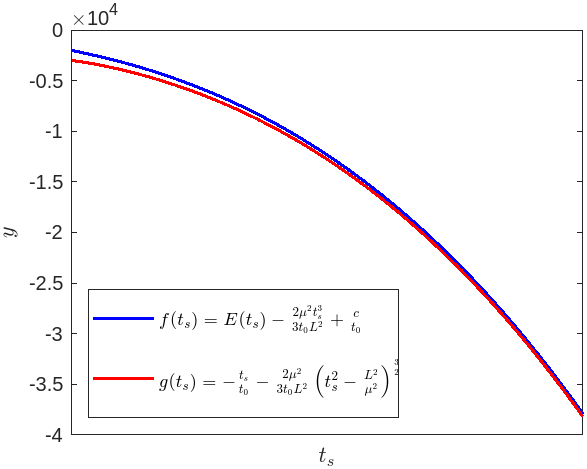}
\caption{$\mu=0.01$, $L=3$, $t_{0}=0.1$ and $c_{9}=0.1$: Blue Colour line for $f(t_{s})$ and Red colour line for $g(t_{s})$ }
\label{fig:15}
\end{figure}

\begin{figure}[h]
\centering            \includegraphics[width=8cm]{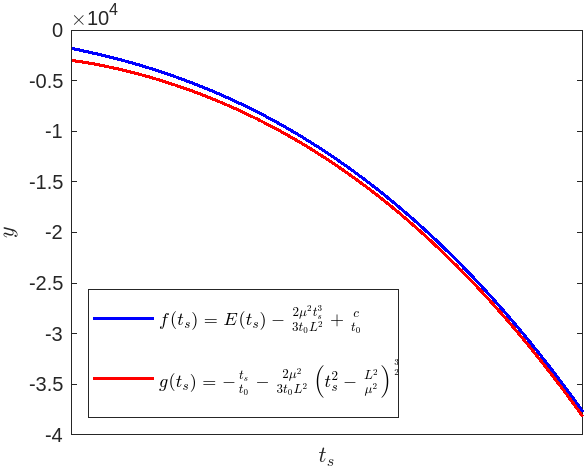}
\caption{$\mu=0.01$, $L=3$, $t_{0}=0.1$ and $c_{9}=20$: Blue Colour line for $f(t_{s})$ and Red colour line for $g(t_{s})$ }
\label{fig:16}
\end{figure}

\begin{figure}[h]
\centering            \includegraphics[width=8cm]{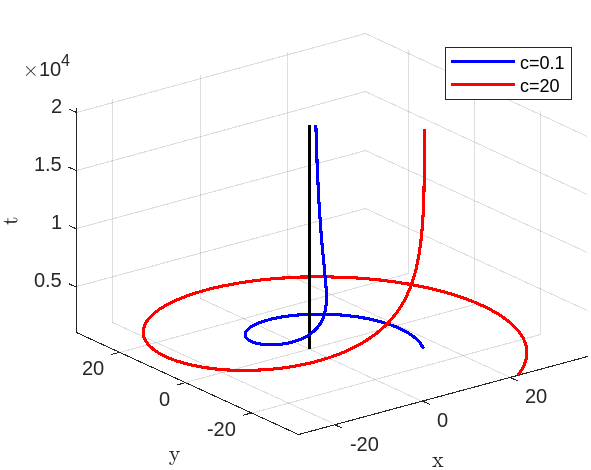}
\caption{Null geodesics for $r-t_{0}E(t) -\bar{A}-c_{9}=0$: When $\mu=0.01$, $L=3$, $t_{0}=0.1$: Blue Colour line for $c_{9}=0.1$ and Red colour line for $c_{9}=20$ }
\label{fig:17}
\end{figure}

\begin{figure}[h]
\centering            \includegraphics[width=8cm]{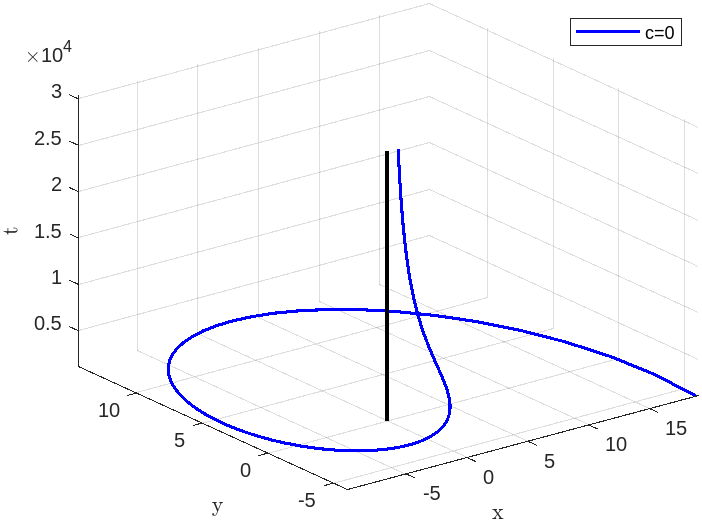}
\caption{Null geodesics for $r-t_{0}E(t) -\bar{A}-c_{9}=0$: When $\mu=0.01$, $L=3$, $t_{0}=0.1$: Blue Colour line for $c_{9}=0$}
\label{fig:21}
\end{figure}

Again, it is also a transcendental equation and it can be analyzed numerically and graphically for any finite value of $\mu$ and $L$. As before, let $f(t_{s})=E(t_{s}) -\frac{2\mu^{2}t^{3}_{s}}{3t_{0}L^{2}}+\frac{c_{9}}{t_{0}}$ and $g(t_{s})=-\frac{t_{s}}{t_{0}}-\frac{2\mu^{2}}{3t_{0}L^{2}}\big(t^{2}_{s}-\frac{L^{2}}{\mu^{2}} \big)^{\frac{3}{2}} $. Then the only possibility for a finite value $t_{s}$ if and only if $f(t_{s}) = g(t_{s})$. \\To start with let us consider $\mu=0.01$ and $L=3$ then $t_{a}>300$. Let $t_{a}=301$ Then from (\ref{132}), $\bar{A}(t_{a})\approx 99.1025$. Now if we consider $t_{0}=0.1$ then for $c_{9}=0.1$, $r_{a}\approx 99.2025$  $\Phi_{a}\approx -33.0675$ and also for $c_{9}=20$, $r_{a}\approx 119.1025$ and $\Phi_{a}\approx -39.7008$ by using (\ref{138}) and (\ref{140}). Figure-(\ref{fig:15}) and Figure-(\ref{fig:16}) shows that the for $c_{9}=0.1$ and $c_{9}=20$, the Eqs. (\ref{143}) do not have solution for finite $t_{s}$ since the the curves $y=f(t_{s})$ and $y=g(t_{s})$ does not intersect each other. From the Eqs. (\ref{134}) and (\ref{135}), we get $r= t_{0}E(t)+\frac{2\mu^{2}}{3L^{2}}\mathcal{O}\big(\frac{1}{t}\big)+c_{9}$ and $\Phi=\frac{1}{L}\big[-\frac{2\mu^{2}}{3L^{2}}\mathcal{O}\big(\frac{1}{t}\big)-c_{9}\big]$ so that for large $t$, we have $r \to c_{9}$ and $\Phi\to -\frac{c_{9}}{L}$. So there is no way to trace any null geodesics for the non-zero value of $c_{9}$ like $c=0.1$ and $c=20$ as shown in Figure-(\ref{fig:17}) using the equations (\ref{140}) and (\ref{141}). But if we take $c_{9}=0$, a particle starts its journey from $r_{a}\approx 99.1025$ and $\Phi_{a}\approx -33.0342$, it finally plunges to the singularity as shown in Figure-(\ref{fig:21}).

Similar to the previous example, if we look at a particle that begins its journey from $r=r_{b}$ in the path described in Eq. (\ref{136}) at time $t=t_{b} > \frac{L}{\mu}$, we obtain the following relations with the same type of meaning from the Eqs. (\ref{136}), (\ref{137}), and (\ref{133}):
\begin{align}
& c_{9}= r_{b}-t_{0}E(t_{b})-\bar{B}(t_{b}) \label{144}\\
& c_{9} \le r_{b}-\bar{B}(t_{b}) < \frac{t_{0}}{2}+c_{9}\label{145}\\
& r-t_{0}E(t) -\bar{B}(t)-r_{b}+t_{0}E(t_{b})+\bar{B}(t_{b})=0 \label{146}\\
& \Phi=\frac{1}{L}\Big[-\bar{B}(t)-r_{b}+t_{0}E(t_{b})+\bar{B}(t_{b})\Big] \label{147}\\
&  \Phi=\Phi_{b}=\frac{1}{L}\Big[-r_{b}+t_{0}E(t_{b})\Big] \label{148}\\
& E(t_{s}) -\frac{2\mu^{2}t^{3}_{s}}{3t_{0}L^{2}}+\frac{c_{9}}{t_{0}}=-\frac{t_{s}}{t_{0}}+\frac{2\mu^{2}}{3t_{0}L^{2}}\Big(t^{2}_{s}-\frac{L^{2}}{\mu^{2}} \Big)^{\frac{3}{2}} \label{149}
\end{align}

\begin{figure}[h]
\centering            \includegraphics[width=8cm]{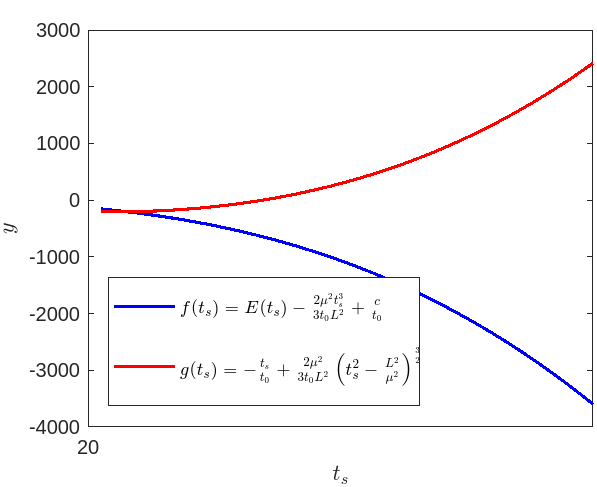}
\caption{$\mu=0.05$, $L=1$, $t_{0}=0.1$ and $c_{9}=0.1$: Blue Colour line for $f(t_{s})$ and Red colour line for $g(t_{s})$ }
\label{fig:18}
\end{figure}

\begin{figure}[h]
\centering            \includegraphics[width=8cm]{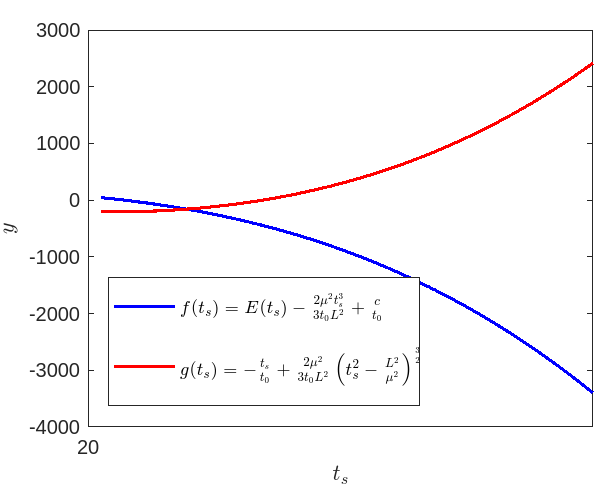}
\caption{$\mu=0.05$, $L=1$, $t_{0}=0.1$ and $c_{9}=20$: Blue Colour line for $f(t_{s})$ and Red colour line for $g(t_{s})$ }
\label{fig:19}
\end{figure}

\begin{figure}[H]
\centering            \includegraphics[width=8cm]{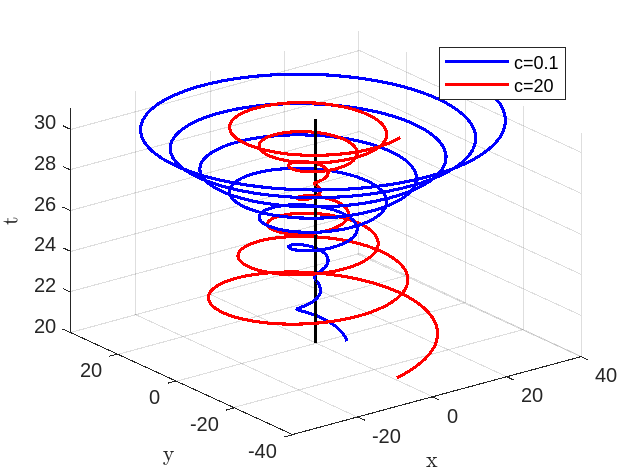}
\caption{Null geodesics for $r-t_{0}E(t) -\bar{B}-c_{9}=0$: When $\mu=0.05$, $L=1$, $t_{0}=0.1$: Blue Colour line for $c_{9}=0.1$ and Red colour line for $c_{9}=20$ }
\label{fig:20}
\end{figure}

Again, as before, Eq. (\ref{149}) is a transcendental equation, so it can be analyzed numerically and graphically for any finite value of $\mu$ and $L$. Now let $f(t_{s})=E(t_{s}) -\frac{2\mu^{2}t^{3}_{s}}{3t_{0}L^{2}}+\frac{c_{9}}{t_{0}}$ and $g(t_{s})=-\frac{t_{s}}{t_{0}}+\frac{2\mu^{2}}{3t_{0}L^{2}}\big(t^{2}_{s}-\frac{L^{2}}{\mu^{2}} \big)^{\frac{3}{2}} $. Then the only possibility for a finite value $t_{s}$ if and only if $f(t_{s}) = g(t_{s})$. \\To start with let us consider $\mu=0.05$ and $L=1$ then $t_{b}>20$. Let $t_{b}=21$, then from (\ref{133}), $\bar{B}(t_{a})\approx 5.1275$. Now if we consider $t_{0}=0.1$ then for $c_{9}=0.1$, $r_{b}\approx 5.2275$  $\Phi_{b}\approx -5.2275$ and also for $c_{9}=20$, $r_{b}\approx 25.1275$ and $\Phi_{b}\approx -25.1275$ by using (\ref{136}) and (\ref{137}). Figure-(\ref{fig:18}) shows that for $c_{9}=0.1$, the Eq. (\ref{149}) has a solution for $t_{s}$ since the the curves $y=f(t_{s})$ and $y=g(t_{s})$ intersect each other. Also Figure-(\ref{fig:19}) shows that for $c_{9}=20$, the Eq. (\ref{149}) has a solution for $t_{s}$ since the curves $y=f(t_{s})$ and $y=g(t_{s})$ intersect each other. So for $c_{9}=0.1$ and $c_{9}=20$, the particle which starts from $r=r_{b}$ at time $t=t_{b}>20$ and it will reach the singularity ($r=0$) at $t=t{s}$ as shown in Figure-(\ref{fig:20}) using the equations (\ref{136}) and (\ref{137}).

\section{Discussion and Conclusion}

In the present investigation we have traced time-like geodesics and null geodesics for two different  K-essence Vaidya Mass functions $\mathcal{M}(t,r)=M+\frac{r}{2}e^{-\frac{t}{t_{0}}}$ and  $\mathcal{M}(t,r)=\mu t+\frac{r}{2}e^{-\frac{t}{t_{0}}}$ on equatorial plane $\theta =\frac{\pi}{2}$ by using Euler-Lagrange equations. For both the cases, we observe that $\mathcal{M}(t,r)$ can be written in terms of energy $E(t)$. We have noticed that for the radial time-like and radial null geodesics for both the K-essence Vaidya masses, a particle having zero angular momentum $(L)$ starting from a finite distance takes a finite time to reach the singularity. For the first K-essence Vaidya mass function, there are two types of families for the time-like case that can be traced one for $M=\sqrt{2}L$ and the other for $M>\sqrt{2}L$ and similarly for the null case one for $M=L$ and the other for $M>L$. But a bit of difference we observed for the second K-essence Vaidya mass function. There does exist only one type of family for the time-like case when $\mu t>\sqrt{2}L$ and for the null case $\mu t>L$. For all the above-mentioned cases, for different finite value of arbitrary constants $c$, we have traced different kinds of orbits for both the generalized K-essence Vaidya mass function and the conclusions as follows:\\

Based on Fig.~(\ref{fig:1}), it may be inferred that a future observer will observe the particle approaching to the singularity within a finite amount of time, as previously stated. According to Figs.~\ref{fig:2} -- \ref{fig:4} it is evident that a future observer can watch the particle reaching at the singularity within a finite period and thereafter departing from it. Based on the information provided in Figs.~\ref{fig:1} -- \ref{fig:4}, for non-radial time-like geodesics of the generalized K-essence Vaidya spacetime (\ref{13}) with the first mass function (\ref{17}) or (\ref{27}), it can be inferred that there is a worry with the presence of a central singularity (where both $r$ and $t$ approach to $0$). It is evident that the central singularity in the generalized Vaidya spacetime is defined by the limits $r\rightarrow 0$ and $t\rightarrow 0$, as shown in ~\cite{husain,maombi,gm4,Joshi6,patil}. However, based on our analysis of the given data presented in Figs.~\ref{fig:1} -- \ref{fig:4}, we may conclude that when $r$ approaches zero, $t$ does not tend to zero. This phenomenon is also observed in the non-radial null geodesics with the same mass function but with varying finite time for different situations, as depicted in Figs.~\ref{fig:5} -- \ref{fig:8}. These phenomena can be explained as follows: a future observer is expected to see the particle approaching to the singularity and then moving away from it for a limited period of time (\ref{51}) or (\ref{62}) or (\ref{67}) (non-radial time-like geodesic) and (\ref{80}) or (\ref{95}) or (\ref{96}) (non-radial null geodesic) in different situations. This observation could potentially indicate the presence of a wormhole during the extreme stages of spacetime, specifically the black hole and white hole, similar to the concept of the Einstein-Rosen bridge ~\cite{ER1935a,ER1935b,FH1962,Ellis1973,Morris1988,Visser1989,Hayward2009,Bronnikov2013,Garattini2019,Mishra2021,Mishra2022,Mustafa2022}. Regarding particles, it is clear that their signature is changed by the singularity.  In this interesting scenario, it seems that the presence of a wormhole allows for the possibility of gravitational collapse resulting in either a naked singularity or a black hole and a white hole in the generalized $K$-essence Vaidya spacetime with the given mass function (\ref{27}). Based on the analysis of collapsing scenarios in the generalized Vaidya spacetime, it has been established that there can be either a naked singularity or a black hole \cite{husain,gm4}. This is supported by the potential presence of a dynamical horizon, as discussed in \cite{Ashtekar1,Ashtekar2,Ashtekar3,Hayward}. On the other hand, Vertogradov \cite{vert1} has also addressed the issue and showed the presence of naked singularities as well as white holes within the framework of the usual generalized Vaidya spacetime in a different setting. However, in our case, the results are a bit different as we have already mentioned that there is the possibility of either a naked singularity or a black hole and a white hole. If we look at all the aforementioned figures with a closer view then it actually reveals the presence of a wormhole situation also in between the black hole and white hole scenarios. But here the conditions being extremal ($r \rightarrow 0$ and $t \neq 0$) we have a membrane-type wormhole instead of a tunnel as has been usually suggested. Alternatively, the previous explanation can also be elucidated as follows: Given that the particle reaches the point at $r=0$ and then escapes from there within a certain period, it may be possible that a quantum tunneling event occurs at the vicinity of the central singularity. Since there is a finite time at the typical singular region at $r=0$, it follows that there is likewise a finite probability of escaping the region at $r=0$. We also have a finite time (\ref{43}) and (\ref{73}) to approach the singularity for a particle for the radial time-like and null geodesic situations.

On the other hand, we have the characteristics of the radial and non-radial time-like or null geodesics structures for the case of the second mass function (\ref{97}). Through an analysis of the radial properties of the time-like and null geodesics, we have determined that the particle ultimately approaches the singularity at $r=0$ within a limited amount of time, as shown by Eqs. (\ref{108}) and (\ref{130}) respectively. To investigate the non-radial geodesics of the specified spacetime (\ref{13}) with the mass function (\ref{97}), we have not found a finite expression for time for a particle to reach the singularity at $r=0$. However, we have obtained a transcendental equation for different possibilities, namely Eqs. (\ref{122}) or (\ref{127}) for time-like geodesics and Eqs. (\ref{143}) or (\ref{149}) for null geodesics. But from these transcendental equations, we have analyzed the non-radial time-like or null geodesics through numerical and graphical analysis. Based on the Figs.~\ref{fig:9} -- \ref{fig:14}, our investigation of non-radial time-like geodesics reveals that when $c_{8}=0.1$, the particle is unable to reach at the singularity at $r=0$ (blue line).  This phenomenon may be explained as follows: The presence of a black hole singularity with an event horizon might result in the spectator being unable to perceive the particle's journey towards the singularity. Based on the aforementioned figures (red line), it has been observed that when the constant $c_{8}$ is set to 30, the particle approaches to the singularity at $r=0$ after a finite amount of time and subsequently moves away from it. Here, we also see the occurrence of a wormhole-like physical phenomenon or quantum tunneling effect near the singularity at $r=0$.

Regarding the non-radial null geodesic characteristics of the aforementioned spacetime and mass function (\ref{97}), we note that the transcendental Eqs. (\ref{143}) and (\ref{149}) exhibit the same kind of solutions as Eq. (\ref{131}). In this situation also, we are yet unable to determine the expression for the time it takes to reach the singularity at $r=0$. After solving the transcendental equations using numerical analysis, we find the graphical solutions through the following figures, i.e. Fig.~\ref{fig:15}, Fig.~\ref{fig:16}, Fig.~\ref{fig:17}, Fig.~\ref{fig:21}, Fig.~\ref{fig:18}, Fig.~\ref{fig:19} and Fig.~\ref{fig:20}. From Fig.~\ref{fig:17}, we have not found any tracing of the non-radial null geodesic structure for two choices of the constants {\it viz.} $c_{9}=0.1$ and $c_{9}=20$ though available for the choice of $c_{9}=0$, from Fig.~\ref{fig:18} -- \ref{fig:21}. No traces of the non-radial null geodesic structure were noticed in the Fig.~\ref{fig:17} for two different values of the constants, namely $c_{9}=0.1$ and $c_{9}=20$. However, by setting $c_{9}=0$, as seen in Fig.~\ref{fig:21}, it is evident that the particle originates from a specific location and ultimately drops into the singularity. Alternatively, we have explored an additional solution to Eq. (\ref{131}) and its associated Eq. (\ref{149}). We have analyzed the non-radial null geodesic structure and depicted it in Fig. \ref{fig:20}. In this scenario, we note that for two specific values of the constants, namely $c_{9}=0.1$ and $c_{9}=20$, the particle begins its trip from a fixed position and eventually reaches the singularity within a limited amount of time. After reaching at the singularity, the particle then escapes from it. This once again indicates the presence of a wormhole or the quantum tunneling effect.\\

Based on the aforementioned discussion, we can conclude that in our specific model, we have investigated the existence of a wormhole during the extreme stages of spacetime, specifically the black hole and white hole, similar to the concept of the Einstein-Rosen bridge or quantum tunneling effect near the central singularity. This investigation was carried out in the comoving frame for the generalized K-essence Vaidya spacetime using two types of K-essence Vaidya mass functions. It is possible to argue that extra interactions between the K-essence scalar field and the usual gravity are the cause of this occurrence.  

Based on the investigation of the present work, especially observations on the Figs.~\ref{fig:2} -- \ref{fig:4}, one may raise the issue of the black holes and baby universes~\cite{Hawking1993} that whether there is any possibilities of such situations under the generalized $K$-essence Vaidya spacetime. This obviously draws special attention and can be considered as a potential topic in the near future.

\section*{Acknowledgments}
The research by M.K. was carried out in Southern Federal University with financial support of the Ministry of Science and Higher Education of the Russian Federation (State contract GZ0110/23-10-IF).
S.R. thanks the Inter-University Centre for Astronomy and Astrophysics (IUCAA), Pune, Government of India for providing visiting associateship and also acknowledges the facility availed under ICARD at CCASS, GLA University, Mathura.

\end{document}